\documentclass[iop,twocolappendix]{emulateapj}

\usepackage{epsfig}
\usepackage{amsmath}

\usepackage{natbib}
\usepackage[backref,breaklinks,colorlinks,citecolor=blue]{hyperref}

\usepackage{comment}
\font\sevenrm=cmr7 scaled 1000
\newcommand{\HeII}{He{\sevenrm II}$\lambda$4686}
\newcommand{\tD}{57482.9}
\newcommand{\tdisc}{57537.4}
\newcommand{\Msun}{$M_\odot$}
\newcommand{\Swift}{\textsl{Swift}}
\newcommand{\restdt}{49}
\begin{document}

\title{Revisiting Optical Tidal Disruption Events with \lowercase{i}PTF16\lowercase{axa}}
\author{T. Hung\altaffilmark{1}, S. Gezari\altaffilmark{1,2}, N. Blagorodnova\altaffilmark{3}, N. Roth\altaffilmark{1,2}, S.B. Cenko\altaffilmark{2,4}, S. R. Kulkarni\altaffilmark{3}, A. Horesh\altaffilmark{5}, I. Arcavi\altaffilmark{6,7,8}, C. McCully\altaffilmark{6,7}, Lin Yan\altaffilmark{9,10}, R. Lunnan\altaffilmark{3}, C. Fremling\altaffilmark{11}, Y. Cao\altaffilmark{3,12}, P. E. Nugent\altaffilmark{13,14}, P. Wozniak\altaffilmark{15}}
\altaffiltext{1}{Department of Astronomy, University of Maryland, College Park, MD 20742, USA}
\altaffiltext{2}{Joint Space-Science Institute, University of Maryland, College Park, MD 20742, USA}
\altaffiltext{3}{Department of Astronomy, California Institute of Technology, Pasadena, CA 91125, USA}
\altaffiltext{4}{NASA Goddard Space Flight Center, Mail Code 661, Greenbelt, MD 20771, USA}

\altaffiltext{5}{Racah Institute of Physics, Hebrew University, Jerusalem, 91904, Israel}

\altaffiltext{6}{Department of Physics, University of California, Santa Barbara,
CA 93106-9530, USA}

\altaffiltext{7}{Las Cumbres Observatory, 6740 Cortona Dr. Suite 102, Goleta, CA, 93111}
\altaffiltext{8}{Einstein Fellow}

\altaffiltext{9}{Caltech Optical Observatories, Cahill Center for Astronomy and Astrophysics, California Institute of Technology, Pasadena, CA 91125, USA}
\altaffiltext{10}{Infrared Processing and Analysis Center, California Institute of Technology, Pasadena, CA 91125, USA}

\altaffiltext{11}{Department of Astronomy, The Oskar Klein Center, Stockholm University, AlbaNova, SE-10691, Stockholm, Sweden}
\altaffiltext{12}{eScience Institute and Astronomy Department, University of
Washington, Seattle, WA 98195}
\altaffiltext{13}{Department of Astronomy, University of California, Berkeley, CA 94720-3411, USA}
\altaffiltext{14}{Lawrence Berkeley National Laboratory, 1 Cyclotron Road, MS 50B-4206, Berkeley, CA 94720, USA}
\altaffiltext{15}{Los Alamos National Laboratory, MS D436, Los Alamos, NM, 87545, USA}

\begin{abstract}
We report the discovery by the intermediate Palomar Transient Factory (iPTF) of a candidate tidal disruption event (TDE) iPTF16axa at $z=0.108$, and present its broadband photometric and spectroscopic evolution from 3 months of follow-up observations with ground-based telescopes and \textsl{Swift}. The light curve is well fitted with a $t^{-5/3}$ decay, and we constrain the rise-time to peak to be $<$49 rest-frame days after disruption, which is rougly consistent with the fallback timescale expected for the $\sim 5\times$10$^{6}$ \Msun{} black hole inferred from the stellar velocity dispersion of the host galaxy. The UV and optical spectral energy distribution (SED) is well described by a constant blackbody temperature of T$\sim$ 3$\times$10$^4$ K over the monitoring period, with an observed peak luminosity of 1.1$\times$10$^{44}$ erg s$^{-1}$. The optical spectra are characterized by a strong blue continuum and broad HeII and H$\alpha$ lines characteristic of TDEs. We compare the photometric and spectroscopic signatures of iPTF16axa with 11 TDE candidates in the literature with well-sampled optical light curves. Based on a single-temperature fit to the optical and near-UV photometry, most of these TDE candidates have peak luminosities confined between log(L [erg s$^{-1}$]) = 43.4-44.4, with constant temperatures of a few $\times 10^{4}$ K during their power-law declines, implying blackbody radii on the order of ten times the tidal disruption radius, that decrease monotonically with time.  For TDE candidates with hydrogen and helium emission, the high helium-to-hydrogen ratios suggest that the emission arises from high-density gas, where nebular arguments break down. We find no correlation between the peak luminosity and the black hole mass, contrary to the expectations for TDEs to have $\dot{M} \propto M_{\rm BH}^{-1/2}$.
\end{abstract}

\section{Introduction}

A tidal disruption event (TDE) occurs when a star passes close enough to a supermassive black hole (SMBH) for the tidal forces of the black hole to exceed the self-gravity of the star, and the star is torn apart by the encounter. For a stellar approach on a nearly-parabolic orbit, about half of the stellar debris will remain bound to the black hole while the other half gains enough energy to escape the gravitational attraction of the black hole. As the bound material returns to pericenter, the material will feed onto the black hole and generate a flare of radiation. The classical solution assuming a uniform mass distribution predicts the mass fallback rate ($\dot{M}$) to follow a $t^{-5/3}$ power law decay \citep{Rees1988,Phinney1989} that can be expressed as $\dot{M}$ = $\dot{M}_{peak}(t/t_{min})^{-5/3}$. The peak mass fallback rate is defined as $\dot{M}_{peak}$ = $\frac{1}{3}\frac{M_\star}{t_{min}}$ while the fallback time $t_{min}$ is proportional to $M_{\rm BH}^{1/2}$ \citep{2011MNRAS.410..359L}.

Although it is shown that in a more detailed calculation the mass fallback rate is determined by the internal structure of the star and even the spin of the black hole, the fallback rate at late times generally approaches the classical $t^{-5/3}$ power law \citep{2009MNRAS.392..332L,Guillochon2013}. Theoretically, the peak mass accretion rate depends on the mass of the black hole as $\dot{M}_{peak} \propto M_{\rm BH}^{-1/2}$ \citep{2011MNRAS.410..359L}. For a smaller black hole with $M_{\rm BH}$ $\lesssim$ 10$^7$ \Msun{}, the initial stage of the accretion is expected to be super-Eddington \citep{Strubbe2009,2011MNRAS.410..359L}. However, it is still unclear how the fallback rate translates to the observed luminosity.

The first few discoveries of TDEs were made in the 1990s in the form of luminous soft X-ray outbursts in quiescent galaxies from the ROSAT survey \citep{1996A&A...309L..35B,1999A&A...343..775K,1995A&A...299L...5G,1995MNRAS.273L..47B,1999A&A...349L..45K,1999A&A...350L..31G,2000A&A...362L..25G}. Several more TDE candidates with similar properties were found in archival searches with the XMM-Newton Slew Survey \citep{2007A&A...462L..49E,2008A&A...489..543E} and Chandra \citep{2013MNRAS.435.1904M,2014ApJ...781...59D}, until the serendipitous discovery of jetted TDE candidates with hard X-ray spectra and super-Eddington luminosities by the \Swift{} satellite \citep{Bloom2011,Burrows2011, Levan2011,Zauderer2011,Cenko2012}.  Most of the recent discoveries of TDEs have been in the UV and optical, exploiting the wide-field UV capabilities of \textsl{GALEX}, and optical synoptic sky surveys such as the Sloan Digital Sky Survey (SDSS), Palomar Transient Factory (PTF), Pan-STARRS1 (PS1), and the All-Sky Automated Survey for Supernovae (ASASSN).  The UV or optically discovered TDE candidates tend to peak in the UV with blackbody temperatures of a few $\times$ 10$^4$K, while the non-jetted X-ray TDE candidates have temperatures in the range of $\sim (0.6-1.0) \times 10^6$K. It is worth mentioning that all of the optically detected TDE candidates are weak or not detected in the X-rays, except ASASSN-14li \citep{Holoien2016}. 

While the temperature of X-ray TDEs is roughly consistent with the theoretical expectations for radiation powered by mass accretion in the TDE debris disk, the discovery of optical TDEs has challenged this simple picture.  Not only do they have a much lower temperature than expected, but the lack of temperature evolution in tandem with the decreasing accretion rate is also in disagreement with thermal radiation from the debris disk. Many studies have tried to resolve this discrepancy by considering several mechanisms that could lead to the observed signature.  For example, the production of an optically thick envelope that radiates at the Eddington limit \citep{1997ApJ...489..573L}, or a strong disk wind or outflow that regulates the accretion rate \citep{Strubbe2009,2015ApJ...805...83M,2016MNRAS.461..948M}. Alternatively, elliptical accretion may cause energy lost to the black hole before circularization, resulting in 1\%-10\% of the bolometric efficiency of a standard accretion disk \citep{2017MNRAS.tmp..118S}.

Another weakness in the classical picture of TDEs is debris circularization, which was assumed to happen immediately when the debris returns to pericenter ($R_p$) \citep{Rees1988}. Recent work by \cite{2015ApJ...804...85S} has shown that orbital energy cannot be dissipated efficiently at r$\sim$ $R_p$ and therefore the circularization process does not happen as quickly as previously thought. Instead, stream-stream collisions are thought to play an important role in producing shocks that convert kinetic energy into thermal energy \citep{1994ApJ...422..508K}. In \cite{2015ApJ...812L..39D}, the extent of apsidal precession that causes different distances of self-intersection of the tidal debris from the supermassive black hole was proposed to explain why there exists two populations of TDE temperatures. Hydrodynamical simulations also suggest that stream-stream collisions may be responsible for the observed UV/optical emission of TDEs \citep{2015ApJ...806..164P,2015ApJ...804...85S,2017MNRAS.464.2816B,2016ApJ...830..125J}. 

The method for photometric selection of TDEs in optical transient surveys was demonstrated in an archival study of the SDSS Stripe 82 Survey by \cite{vanVelzen2011}, and resulted in the recovery of two likely TDE candidates. Since then, on the order of a dozen of optical TDEs have been discovered promptly enough for spectroscopic follow-up observations, and they show a diversity of broad hydrogen and helium emission line strengths. For example, the optical spectra of PS1-10jh, PTF09ge, and ASASSN-15oi display broad \HeII{}
emission lines with no sign of H$\alpha$ emission, ASASSN-14li
shows both broad prominent HeII and H$\alpha$ emission, and ASASSN-14ae has strong H$\alpha$ emission and a weaker but broad \HeII{} that developed later in time. The spectral family of TDEs was first discussed in \citet{Arcavi2014}. The mechanisms behind the spectroscopic signatures are still under debate. Proposed explanations include the chemical composition of the progenitor star \citep{Gezari2012}, and photoionization conditions in the debris disk \citep{2014ApJ...783...23G} or an optically-thick reprocessing envelope \citep{Roth2016}.

The paper is structured as follows. In \S\ref{sec:discovery} we present the discovery of a newly discovered optical TDE candidate iPTF16axa.  We describe the pre-event data associated with its host galaxy in \S\ref{sec:archival data} and the follow up photometric and spectroscopic observations we obtained for iPTF16axa in \S\ref{sec:followup obs}. The results of SED and spectral analyses are presented in \S\ref{sec:analysis}. In \S\ref{sec:discussion}, we compare the physical quantities derived from the SEDs and the spectral measurements with 11 UV/optical events that are classified as strong TDE candidates with well-sampled optical light curves.

\section{Discovery of \lowercase{i}PTF\lowercase{16axa}}
\label{sec:discovery}

iPTF16axa (right ascension, $\alpha_{J2000}$ = 17h03m34.36s; declination, $\sigma_{J2000}$ = +30$^{\circ}$35'36.8") is a TDE discovered by the intermediate Palomar Transient Factory (iPTF) using the Palomar 48-inch (P48) telescope. The flare was first detected on UT 2016 May 29 (UT dates are use throughout the paper) with a host flux subtracted magnitude of $g$ =19.49 $\pm$ 0.07 mag. Astrometrically aligned P48 images show that the position of the flare is coincident with the nucleus of the host galaxy, with an offset of 0.17 arcsec that is within the positional uncertainties measured for a reference AGN sample of 0.3 arcsec. Constraints on the peak time are not available since the field is not regularly monitored by iPTF. However, the PTF survey visited this field in 2011 Mar-Sep, 2012 Mar, 2013 Aug, and 2014 May-Jun. No historical variability activity was detected to a 3$\sigma$ limiting magnitude of R~$\approx$~21 mag in any observations during the aforementioned period, which indicates that the source of the flare is unlikely to be caused by a variable active galactic nucleus (AGN).

We requested target-of-opportunity (ToO) observation of iPTF16axa on 2016 June 01 using our Cycle 12 \Swift{} key project (PI Gezari) triggers that are designed for a systematic follow up of iPTF nuclear transients with red host galaxies. The transient satisfies our selection criteria:  observations made with the Palomar-60 inch (P60) telescope shows the transient has a blue color ($g-r \sim -0.4$ mag) and is found in a red host galaxy ($u-g$=1.94 mag and $g-r$=0.91 mag) as is revealed by the Sloan Digital Sky Survey Data Release 9 (SDSS DR9). 

The observation made with the \Swift{} satellite on Jun 07 2016 using the UV-optical telescope (UVOT) in the \textsl{uvw2} filter showed signs of a UV bright source.  After triggering the Swift ToO observation, a classification spectrum was also taken with Keck DEIMOS on 2016 June 04. The classification spectrum shows conspicuous broad \HeII{} line as well as H$\alpha$ emission lines at $z=0.108$ that are indicative of TDEs discovered in the optical. However, a simultaneous \Swift\ X-Ray Telescope (XRT) observation did not show any sign of X-ray emission in 0.3--10keV. A VLA observation made on June 12 also resulted in null detection with a rms of 13 $\mu$Jy at 6.1 GHz and 15 $\mu$Jy at 22 GHz. With the Swift UVOT photometry and the classification spectrum confirming iPTF16axa being a strong TDE candidate, we triggered a series of follow up programs in the UV and optical over a span of three months until the target was not observable by ground-based telescopes.

\section{Archival data}
\label{sec:archival data}
The celestial position of iPTF16axa was covered by SDSS. A galaxy associated with this position, SDSS J170334.34+303536.6, has photometry measurements in $ugriz$. However, no spectroscopy is found to be associated with the host galaxy. 

Archival AllWISE data \citep{2013yCat.2328....0C} shows that the host galaxy was observed with 15.8 mag, 15.5 mag, $<$11.7 mag, and $<$9.4 mag in 3.4, 4.6, 12, and 22 $\mu$m, respectively, in the Vega system. Pre-event GALEX or ROSAT limits of the host galaxy are not available.

\subsection {Host galaxy properties}
\label{subsec:host_prop}
Due to the long-lived nature of TDE, we were not able obtain a host galaxy spectrum of iPTF16axa for this analysis before it went behind the Sun in October. 

We perform synthetic stellar population template fitting to the SDSS broadband photometry \texttt{cmodelMag} in $ugriz$ as well as the WISE 3.4 $\mu$m and 4.6 $\mu$m photometry with Fitting and Assessment of Synthetic Templates (\texttt{FAST}) by \cite{2009ApJ...700..221K}. Assuming an exponentially declining star formation history with the \cite{2003MNRAS.344.1000B} templates, a Salpeter IMF, the \cite{1989ApJ...345..245C} dust extinction law, and $A_V$=0.12 from the \cite{2011ApJ...737..103S} dust map, the fitting program yields a ${\chi_\nu^2}$ of 1.57. The results of the fit suggest that star formation has quenched in the galaxy with an SFR of 10$^{-6.6}$ \Msun{} yr$^{-1}$. 
In February 2017, we obtained a high-resolution spectrum of iPTF16axa with the Echellette Spectrograph and Imager (ESI) mounted on the Keck-II telescope (PI Gezari). We observed the host galaxy and a template GIII star BD+332423 with the 0.5" slit for a total integration time of 3600s and 120s, respectively. The data is reduced with the MAuna Kea Echelle Extraction (\texttt{MAKEE} \footnote{http://www.astro.caltech.edu/~tb/makee/}) package while the wavelength is calibrated with \texttt{IRAF}.

\begin{figure}
\centering
\includegraphics[width=3.5in, angle=0]{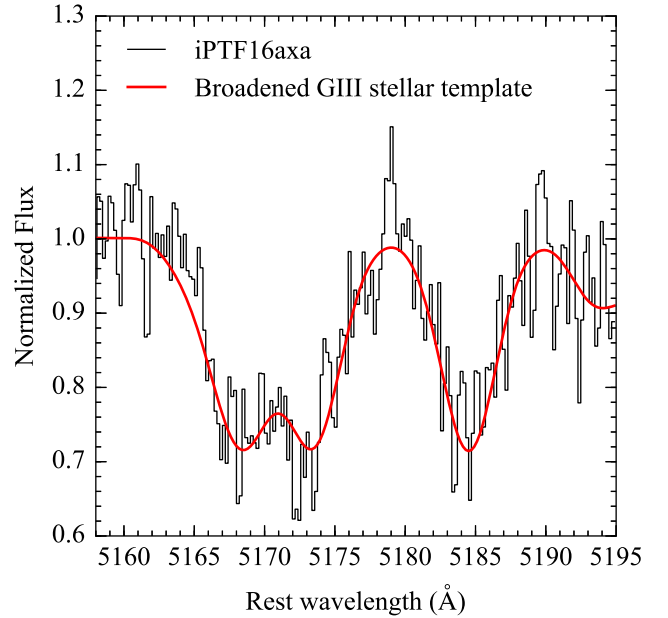}
\caption{Spectral fit around the Mg I$b$ triplet region. The black line shows the Keck ESI spectrum of the host galaxy of iPTF16axa. The red line marks the Keck ESI spectrum of a GIII star BD+332423 that has been broadened to fit the absorption linewidths in the host.}
\label{fig:mgdisp}
\end{figure}

We measure a stellar velocity dispersion of 101.3$\pm$1.9 kms$^{-1}$ with the Mg I$b$ $\lambda\lambda$5167, 5173, 5184 triplet (\autoref{fig:mgdisp}) by broadening the GIII stellar template to match the linewidths in the host spectrum. The velocity dispersion translates to a black hole mass of 5.0$\substack{+7.0 \\ -2.9}\times$ 10$^{6}$ \Msun{} \citep{2013ApJ...764..184M}. Despite the large intrinsic scatter in the M-$\sigma$ relation (0.38 dex), the black hole mass estimated from velocity dispersion is within the range of allowable black hole masses able to disrupt a solar-type star outside of its event horizon.

\section{Follow-up observations}
\label{sec:followup obs}
\subsection{Photometry}
\subsubsection{P48 and P60 photometry}
On 2016 May 29, the transient iPTF16axa was discovered in the $g$ band while iPTF conducted a seasonal experiment that searches for young supernovae using the P48 telescope in Mould-R and SDSS-g' filters to a depth of $\sim$20.5 mag with a 4 day cadence. The nightly P48 raw images are detrended and astrometrically and photometrically calibrated at the Infrared Processing and Analysis Center (IPAC) \citep{2014PASP..126..674L}. Following the discovery of the transient, we requested a series of observations of the source in $gri$ bands with the robotic Palomar 60-inch telescope in order to keep track of the color evolution. The collected data are processed by the Fremling Automated Pipline \citep{2016A&A...593A..68F} that performs image subtraction with respect to the SDSS images and extracts the PSF magnitude of the source.

\subsubsection{LCO photometry}
We obtained 6 epochs of the Las Cumbres Observatory (LCO) follow-up photometry in $gri$ bands. Host flux subtraction is performed using SDSS references. The LCO light curves are consistent with the P60 data but have larger error bars due to cross subtractions of LCO and SDSS. Therefore, the LCO data are not included in the light curve fit in \autoref{subsec:lightcurves}. The LCO subtractions may be improved once the LCO references are obtained.

\subsubsection{\Swift{} UVOT and XRT photometry}
Following the discovery of TDE emission signatures from the spectroscopy, we requested and were granted 13 target-of-opportunity observations spanning a time period of $\sim$ 2.5 months with \Swift{}. The observations were made in all 6 filters of UVOT: UVW2 (1928 \AA), UVM2	(2246 \AA), UVW1 (2600 \AA), U (3465 \AA), B (4392 \AA), and V (5468 \AA). We used a 5" radius aperture and a 20" background region to extract the photometry of the UV source with the task \texttt{uvotsource} in HEASoft \footnote{\url{https://heasarc.gsfc.nasa.gov/lheasoft/}}. Note that due to the lack of pre-event UV limits, we do not attempt to perform host subtraction with the UVOT images.

We also observed the location of iPTF16axa with the XRT \citep{bhn+05} on-board the \Swift{} satellite \cite{gcg+04}
beginning at 4:32 UT on 7 June 2016. Regular monitoring of the
field in photon counting (PC) mode continued for the next 10 weeks.

No significant emission is detected in individual epochs. 
Using standard XRT analysis procedures (e.g., \citealt{ebp+09}), we 
place 90\% confidence upper limits ranging from $(2.9$--$12.0) 
\times 10^{-3}$ counts s$^{-1}$ in the 0.3--10.0\,keV bandpass 
over this time period.  Stacking all the XRT data obtained over 
this period together (29\,ks of total exposure time) also results
in an upper limit of $2.7 \times 10^{-4}$ counts s$^{-1}$.

To convert this count rate to a flux, we adopt a power-law spectrum
with a photon index of $\Gamma = 2$ and incorporate absorption from
the Milky Way (but none in the TDE host galaxy).  We then find
an upper limit on the time-averaged unabsorbed X-ray flux from the location
of iPTF16axa of $< 1.1 \times 10^{-14}$\,erg\,cm$^{-2}$\,s$^{-1}$
(90\% confidence limit).
At the distance of iPTF16axa, this corresponds to a 0.3--10.0\,keV
X-ray luminosity of $L_{X} < 3.3 \times 10^{41}$\,erg\,s$^{-1}$.
While this limit is significantly fainter than the luminous X-ray
emission observed from ASASSN-14li \citep{Holoien2016,vanVelzen2016}, it
is comparable to the much fainter emission observed from
ASASSN-15oi ($L_{X} = 4.8 \times 10^{41}$\,erg\,s$^{-1}$;
\citealt{Holoien2016b}).  And it is several orders of magnitude above
the faint X-ray emission observed at the location of iPTF16fnl
($L_{X} = 2.4 \times 10^{39}$\,erg\,s$^{-1}$; \cite{2017arXiv170300965B}).

\subsection{Spectroscopy}
\subsubsection{Keck DEIMOS}
A Keck DEIMOS classification spectrum was scheduled 3 days after the first \Swift{} ToO observation was triggered (Jun 04 2016). The spectrum was taken with a 0.8" wide slit along with the LVMslitC slit mask and a 600ZD grating. The on-source exposure time was 360s. The data was reduced using the DEIMOS DEEP2 data reduction pipeline with flux calibrated by the spectrum of a spectrophotometric standard star, BD+28d4211, taken on the same night.
\subsubsection{Keck LRIS}
Keck LRIS spectra were taken on Jun 10 2016 and Jul 06 2016. Same configuration were used and the integration time was 900s in both nights. The spectra were taken with a 1" slit and a 400/3400 grism that yields a FWHM resolution of $\sim$7 \AA. The data were reduced with the LRIS automated reduction pipeline \footnote{\url{http://www.astro.caltech.edu/~dperley/programs/lpipe.html}}. The observed flux standard star is BD+28d4211.   
\subsubsection{DCT DeVeny}
An exposure of iPTF16axa was taken on Jun 13 with the 4.3-meter DeVeny spectrograph mounted on the Discovery Channel Telescope (DCT). A 1.5" slit and a 300g/mm grating were used with a central wavelength setting of 5800 \AA. The spectral coverage is 3600-8000 \AA~ at a dispersion of $\sim$2.2 \AA \ per pixel, yielding a FWHM resolution of $\sim$9 \AA. Data were reduced with standard IRAF routines, which include bias removal, flat-fielding, 1-d spectrum extraction, wavelength calibration and flux calibration using spectrophotometric standard star BD+40d4032.

\section{ANALYSIS}
\label{sec:analysis}
Throughout this paper, we correct for Galactic extinction for all data used for analysis using the \cite{1989ApJ...345..245C} extinction curve with $R_C=3.1$ and $E(B-V) = 0.0390$ based on \cite{2011ApJ...737..103S} dust map. We use a luminosity distance of $d_L = 505$ Mpc based on a WMAP9 cosmology with H$_0$ = 69.32 km s$^{-1}$ Mpc$^{-1}$ , $\Omega_M = 0.29$, $\Omega_\Lambda$ = 0.71.

\subsection{Light curves}
\label{subsec:lightcurves}
\begin{figure}
\centering
\includegraphics[width=3.5in, angle=0]{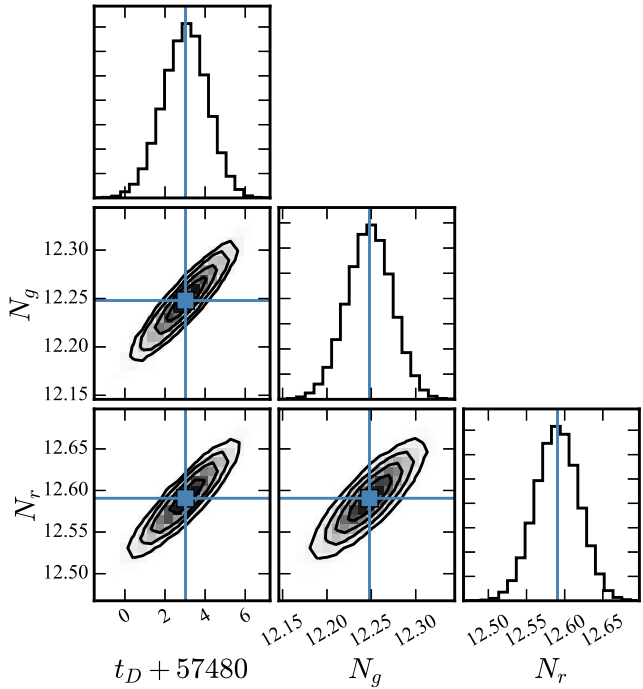}
\caption{Corner plot of the light curve fit, which contains 1000 MCMC simulations.}
\label{fig:corner}
\end{figure}

Classical calculations assume a uniform distribution in specific energy so that the bound stellar debris returns to the pericenter at a rate of $t^{-5/3}$ \citep{Rees1988,Phinney1989}. For more realistic energy distributions, there are deviations from $t^{-5/3}$ at early times, but the light-curve eventually approaches a $t^{-5/3}$ power-law at late times \citep{2009MNRAS.392..332L}, or approaches a power-law index within a range of values that brackets -5/3 \citep{Guillochon2013}. The $t^{-5/3}$ power-law can be expressed as
\begin{equation} 
L(t) \propto \dot{M}(t) \propto (t-t_D)^{-5/3},
\label{eq:powerlaw}
\end{equation}
where $t_D$ is the time of disruption.
We fit the light curves in both $g$ and $r$ bands simultaneously with data taken by P48 and P60 telescopes. With a fixed power-law index of -5/3, we can rewrite \autoref{eq:powerlaw} to
\begin{equation}
m_{obs} = N + \frac{5}{2} \cdot \frac{5}{3}(t-t_D),
\end{equation}

where $m_{obs}$ is the observed magnitude and $N$ is a normalization constant.
We derived a disruption time ($t_D$) of MJD \tD{}$\pm$1.1 using \texttt{emcee} \citep{emcee}, a python implementation of the Affine invariant Markov chain Monte Carlo (MCMC) ensemble sampler. The corner plot for 1000 MCMC simulations is shown in \autoref{fig:corner}. If we loosen the fitting parameter constraints further by allowing the power law index to change freely, we obtain a best-fit power-law index of -1.44$^{+0.09}_{-0.12}$ and a $t_D$ of 57494.7$\pm$0.1. The derived values imply the rise time to peak light is shorter than \restdt{} rest-frame days assuming the peak was reached some time before the discovery of iPTF16axa.

\begin{figure}
\centering
\includegraphics[width=3.5in, angle=0]{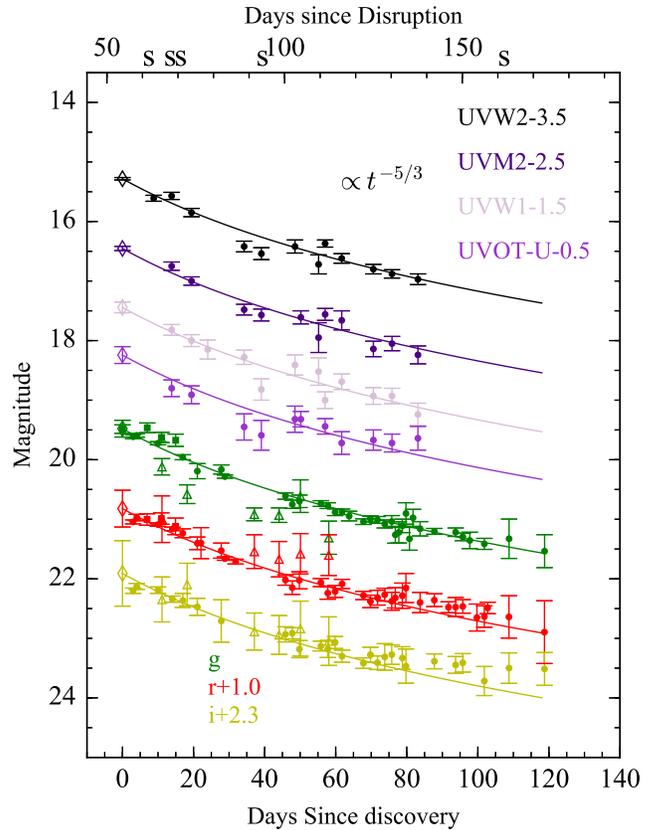}
\caption{The light curve of iPTF16axa with a $t^{-5/3}$ power law fit and dates normalized to the derived disruption time MJD\tD{}. The circles and squares for $gri$bands denote the host-subtracted data taken with P60 and P48 respectively while the diamonds are the extrapolated magnitudes at the time of iPTF discovery. The open triangles in $gri$ bands mark the LCO host-subtracted magnitudes. Note that the LCO data are not included in the light curve fit since they the cross subtractions of LCO data and SDSS result in larger error bars.}
\label{fig:lc}
\end{figure}

The light curve is fitted well by a $t^{-5/3}$ power-law in all the UV and optical bands, with a constant color between the bands with time. The model fit shown in \autoref{fig:lc} has the colors $UVW2-r = -1.05$ mag and $g-r = -0.34$. The lack of color evolution and the observed $t^{-5/3}$ power-law decline in the UV and optical bands requires a fixed temperature over time to be consistent with the expected $t^{-5/3}$ evolution of the bolometric luminosity.

\begin{figure}
\centering
\includegraphics[width=3.5in, angle=0]{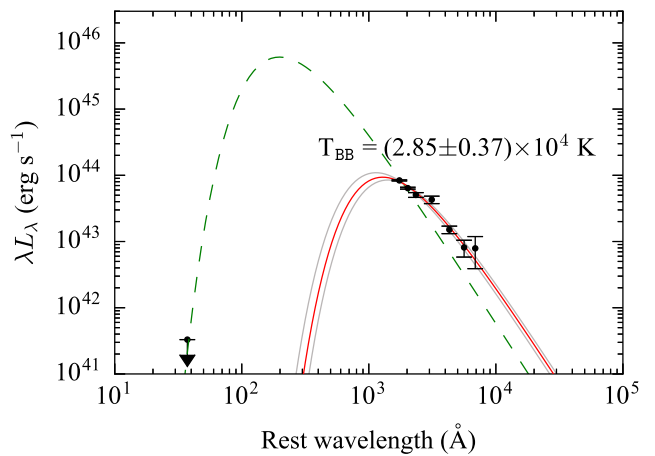}
\caption{Blackbody fit of the UV-optical SED derived from the $t^{-5/3}$ power-law fit in \autoref{subsec:lightcurves} extrapolated to $t_{disc}$= MJD \tdisc{}. The grey curves show the highest and lowest temperatures bounded by the 90\% confidence interval. The green curve shows the blackbody spectrum corresponding to $T_{bb}$=1.85$\times$10$^5$K, which is an upper limit on the temperature imposed by the stacked \Swift{} XRT flux in 0.3keV-10keV.}
\label{fig:BBat_t0}
\end{figure}

In \autoref{fig:BBat_t0}, we show the best-fit blackbody spectrum implied by these colors using the magnitudes extrapolated from the power-law in \autoref{fig:lc} to the time of discovery MJD \tdisc{} in $UVW2$, $UVM2$, $UVW1$, $u$, $g$, $r$, and $i$ bands. The best-fit blackbody temperature implied by the light curve model is 2.85 $\times$ 10$^4$ K. Using the X-ray upper limit, we also place an upper limit of 1.85 $\times$ 10$^5$ K on the blackbody temperature of the TDE, which is shown in the green dashed line in \autoref{fig:BBat_t0}.

\subsection{SED analysis}
Given the archival SDSS $u$ band magnitude of the host is $\sim$ 21.4 mag, we assume the host light contribution is negligible in UVOT filters with shorter wavelengths ($UVW2, UVM2, UVW1, u$). The data in $B$ and $V$ bands are excluded for data analysis since the contribution from the host galaxy is unknown. We also collected the host-subtracted photometry in $gri$ from P48 and P60, which use images from IPAC and SDSS as references, respectively. 

To construct UV-optical SEDs at different epochs, we interpolate the host-subtracted flux in $g$ and $r$ bands using the $t^{-5/3}$ light curves in \autoref{fig:lc} to the epoch of \Swift{} observations. The uncertainties of the interpolated magnitudes in $g$ and $r$ are estimated to be the weighted residual of the light curve fit with the $t^{-5/3}$ power law. 

\begin{figure*}
\centering
\includegraphics[width=7in, angle=0]{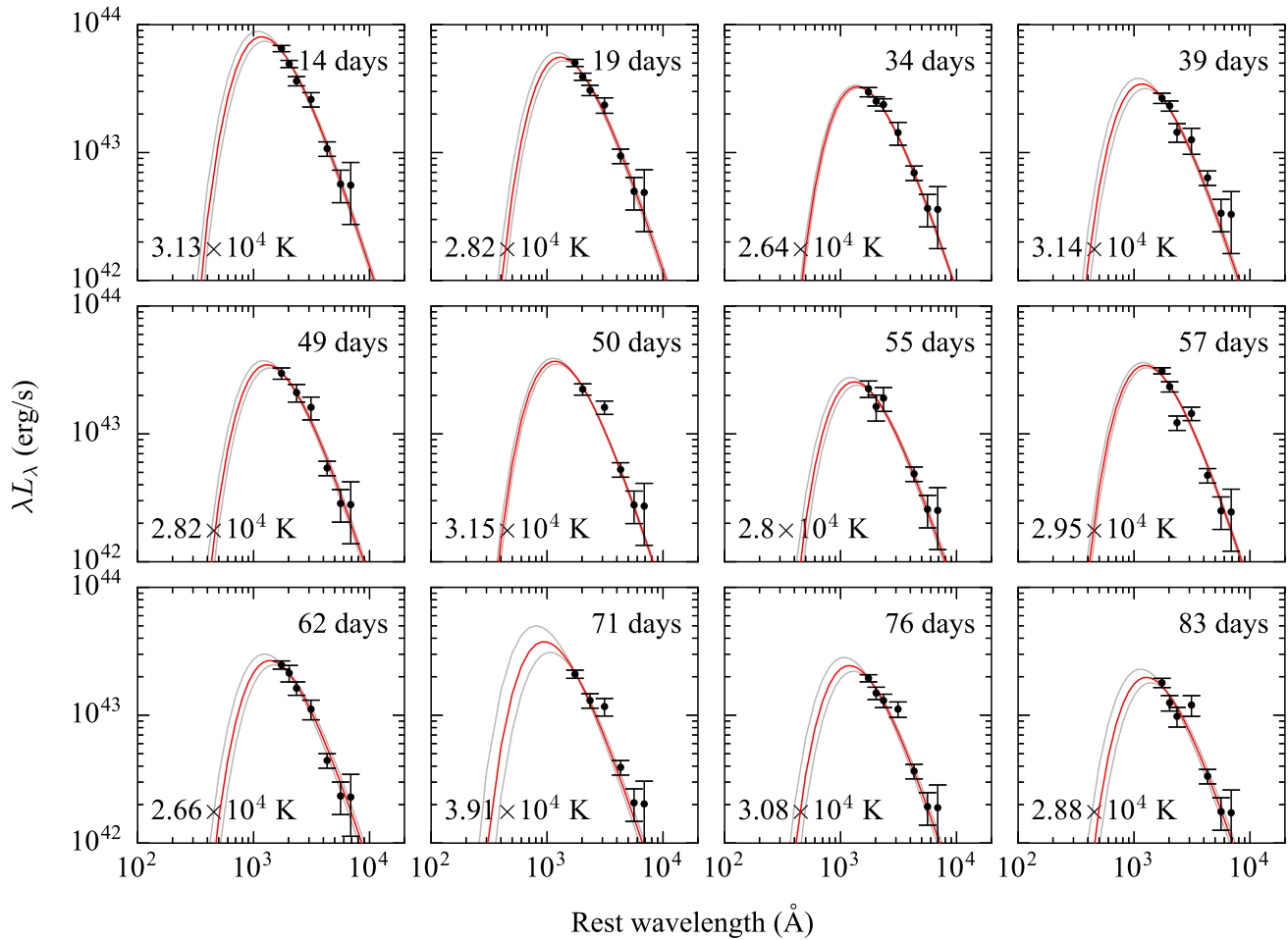}
\caption{The blackbody fit for the transient SEDs. The time indicated in each panel shows the time elapsed since discovery $t_{disc}$ = MJD \tdisc{}. The blackbody temperature remains roughly constant with a mean temperature 3$\times$10$^4$K over time.}
\label{fig:bbfit}
\end{figure*}

\begin{figure}
\centering
\includegraphics[width=3.5in, angle=0]{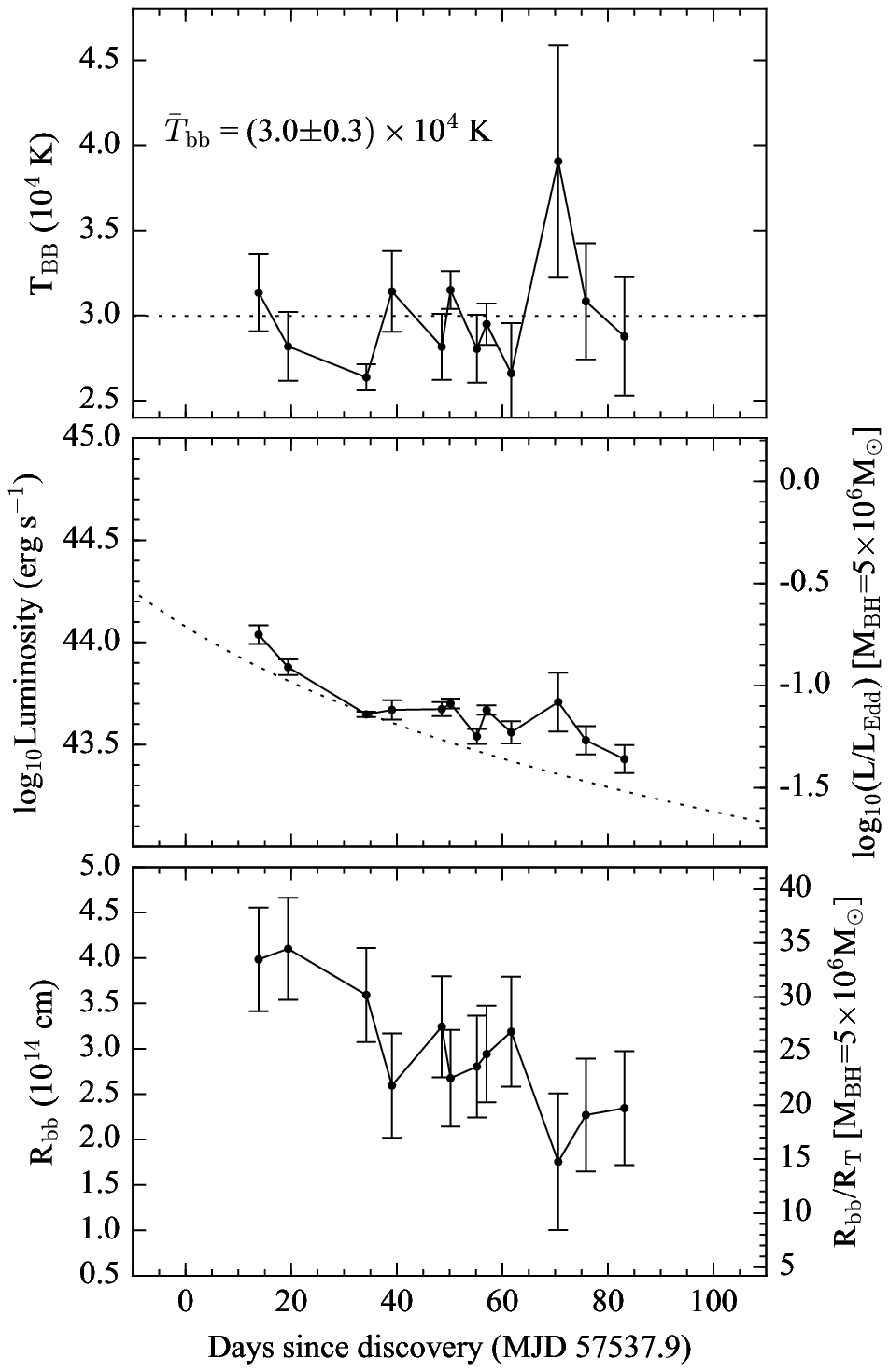}
\caption{The time evolution of iPTF16axa. Upper: The time evolution of blackbody temperature of iPTF16axa. The black dotted line marks the mean $T_{bb}$ of 3$\times$10$^4$ K. Middle: The evolution of integrated UV-optical luminosity. The black dotted line shows the $t^{-5/3}$ prediction from the light curve with a peak luminosity indicated in \autoref{fig:BBat_t0}. Total power emitted (Area under the dotted line integrated from $t_{disc}$ to $t_\infty$) is 5.5$\times$10$^{50}$ ergs. Lower: The time evolution of the blackbody radius inferred from SED fitting. }
\label{fig:evo}
\end{figure}

The SEDs are fit with a blackbody using Markov chain Monte Carlo (MCMC) method. Shown in \autoref{fig:bbfit} are the best-fit blackbody spectra. The 1$\sigma$ uncertainties of the model parameter are shown by the two grey lines in each panel, representing the upper and the lower bound of the best-fit temperature. The best-fit blackbody temperatures are plotted as a function of time in the top panel of \autoref{fig:evo}. The blackbody temperatures of iPTF16axa remained nearly constant temperature $\bar{T}$=(3.0$\pm$0.33)$\times$10$^4$K over 80 days of the \Swift{} monitoring.

In \autoref{fig:evo}, we also plot the time evolution of the UV-optical integrated luminosity and the blackbody radius in the middle and the bottom panels.
We calculate the luminosity by integrating the area under the best-fit blackbody for each SED, which follows the theoretical $t^{-5/3}$ law. Given there is no detection in the X-ray, we assume the bolometric luminosity of the transient is dominated by the emission in the UV and optical. The observed peak luminosity of $1.1 \times 10^{44}$ erg s$^{-1}$ corresponds to an Eddington ratio of 17.4\% for a 5$\times$10$^{6}$ \Msun{} black hole and a mass accretion rate ($\dot{M}_0$) of 1.8$\times$10$^{-2}$~($\epsilon$/0.1)$^{-1}$ \Msun{} yr$^{-1}$, where $\epsilon$ is the accretion efficiency. The total energy integrated under the model fit from from $t_{disc}$ to t$_\infty$ is 5.5$\times$10$^{50}$ erg, which corresponds to a total mass accreted of 3.1$\times$10$^{-3}$  ($\epsilon$/0.1)$^{-1}$ \Msun{}. Note, that this is a small fraction of the $0.5 M_\star$ of mass expected to remain bound to the black hole in a TDE \citep{Rees1988} unless the radiative efficiency ($\epsilon$) is low .

We also calculate the emitting radius of the blackbody using the Stefan-Boltzmann law:

\begin{equation}
L = 4\pi R_{bb}^2 \sigma T_e^4, 
\end{equation}
where L is the luminosity integrated from the best-fit blackbody spectrum to the SED, $R_{bb}$ is the blackbody radius, and $T_e$ is the effective temperature, which is set to be equal to the blackbody temperature derived from the SED fit. In \autoref{table:bbfitparams} we list our fits for the blackbody temperature, luminosity, and photospheric radius for each of our photometric observations. In the bottom panel of \autoref{fig:evo}, we plot our fit for the photospheric radius as a function of time, where the y-axis on the right hand side of \autoref{fig:evo} shows the radius in units of the tidal radius ($R_T$) assuming the disrupted star is a solar mass star. The tidal radius, $R_T = R_\star (M_{\rm BH}/M_\star)^{1/3}$, is 1.19$\times$10$^{13}$~cm for a 5$\times$10$^{6}$\Msun{} black hole.

\begin{deluxetable}{lcccc}

\tablecolumns{5}
\tablecaption{Blackbody fitting from light curves \label{table:bbfitparams}} 
\centering
\tablehead{\colhead{MJD} & \colhead{$t-t_{disc}$} & \colhead{BB Temperature} & \colhead{BB Radius} & \colhead{Luminosity} \\
\colhead{} & \colhead{days} & \colhead{10$^4$ K} & \colhead{10$^{14}$ cm} & \colhead{10$^{43} $erg s$^{-1}$ }
}
\startdata 

57551.23	&	14	&	3.13$\pm$0.23	&	3.98$\pm$2.27	&	10.91$\pm$1.14 \\
57556.82	&	19	&	2.82$\pm$0.2	&	4.1$\pm$2.31	&	7.56$\pm$0.66\\
57571.63	&	34	&	2.64$\pm$0.08	&	3.59$\pm$1.86	&	4.44$\pm$0.13\\
57576.48	&	39	&	3.14$\pm$0.24	&	2.6$\pm$1.49	&	4.67$\pm$0.51\\
57585.93	&	49	&	2.82$\pm$0.19	&	3.24$\pm$1.81	&	4.71$\pm$0.37\\
57587.59	&	50	&	3.15$\pm$0.11	&	2.68$\pm$1.42	&	5.02$\pm$0.28\\
57592.58	&	55	&	2.8$\pm$0.2	&	2.8$\pm$1.57	&	3.46$\pm$0.29\\
57594.42	&	57	&	2.95$\pm$0.12	&	2.94$\pm$1.57	&	4.67$\pm$0.25 \\
57599.09	&	62	&	2.66$\pm$0.29	&	3.19$\pm$1.93	&	3.63$\pm$0.45\\
57608.0	&	71	&	3.91$\pm$0.68	&	1.76$\pm$1.32	&	5.11$\pm$1.69 \\
57613.24	&	76	&	3.08$\pm$0.34	&	2.27$\pm$1.41	&	3.32$\pm$0.53 \\
57620.54	&	83	&	2.88$\pm$0.35	&	2.35$\pm$1.47	&	2.68$\pm$0.43
\enddata

\end{deluxetable}

\subsection{Spectral analysis}
\label{subsec:spectral}
\begin{figure}
\centering
\includegraphics[width=3.5in, angle=0]{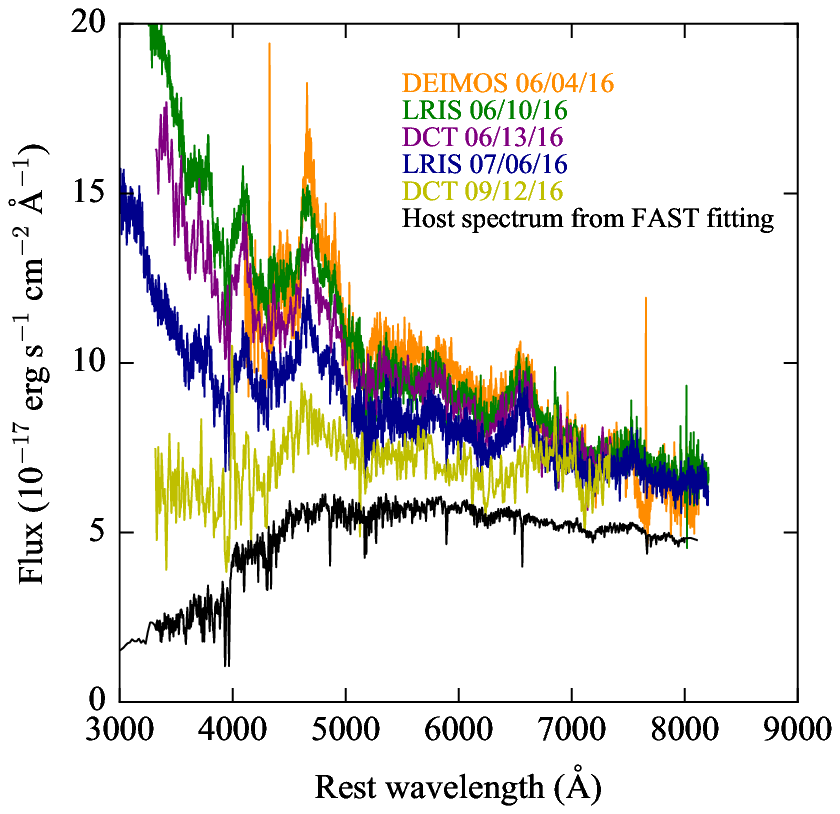}
\caption{Newly observed spectra and host spectrum obtained from fitting the SDSS broadband photometry.}
\label{fig:allspec}
\end{figure}

Five follow up spectra are shown in \autoref{fig:allspec}. We also show the best-fit host spectrum from \texttt{FAST} fit with SDSS \texttt{fiberMag} (flux enclosed in a 3" diameter fiber) in $ugriz$ filters as described in \autoref{subsec:host_prop}. In \autoref{fig:allspec}, we rescale all the new spectra to the synthetic magnitude in the $r$ band ($m_{r,syn}$), which is defined as

\begin{equation}
m_{r,syn} = -2.5\textrm{log}_{10}(10^{-m_{r,0}/2.5} + 10^{-m_{r,sub}/2.5}),
\end{equation}

where $m_{r,0}$ is the fiber magnitude of the host and $m_{r,sub}$ is the host-subtracted $r$ band magnitude derived from the $t^{-5/3}$ power-law fit at the time of the observation.

\begin{figure*}
\centering
\includegraphics[width=7.0in, angle=0]{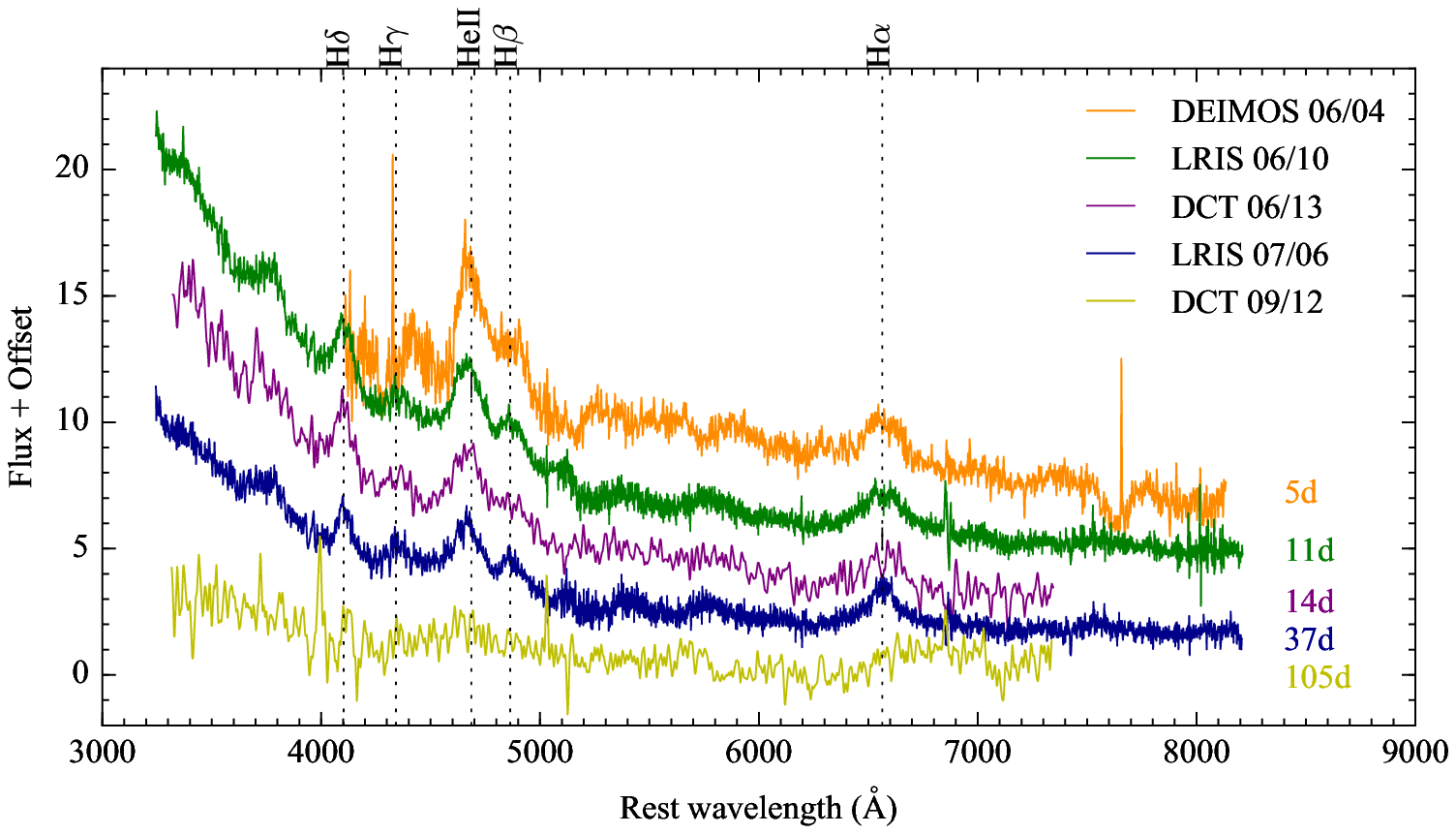}
\caption{Host-subtracted spectra of the new observations. The flux levels are offset for better visualization. The DEIMOS spectrum is smoothed by two pixels.}
\label{fig:allspec_sub}
\end{figure*}

In order to measure the broad emission lines in the spectra, we first subtract off the instrumental broadened host template spectrum from the \texttt{FAST} fit in \autoref{subsec:host_prop}, where $\sigma$ = $\sqrt{\sigma_{instrument}^2 - \sigma_{lib}^2}$, for each spectrum. The FWHM resolution of the \cite{2003MNRAS.344.1000B} template library is 3\AA. The host-subtracted spectra are shown in \autoref{fig:allspec_sub}.

\begin{figure}
\centering
\includegraphics[width=3.5in, angle=0]{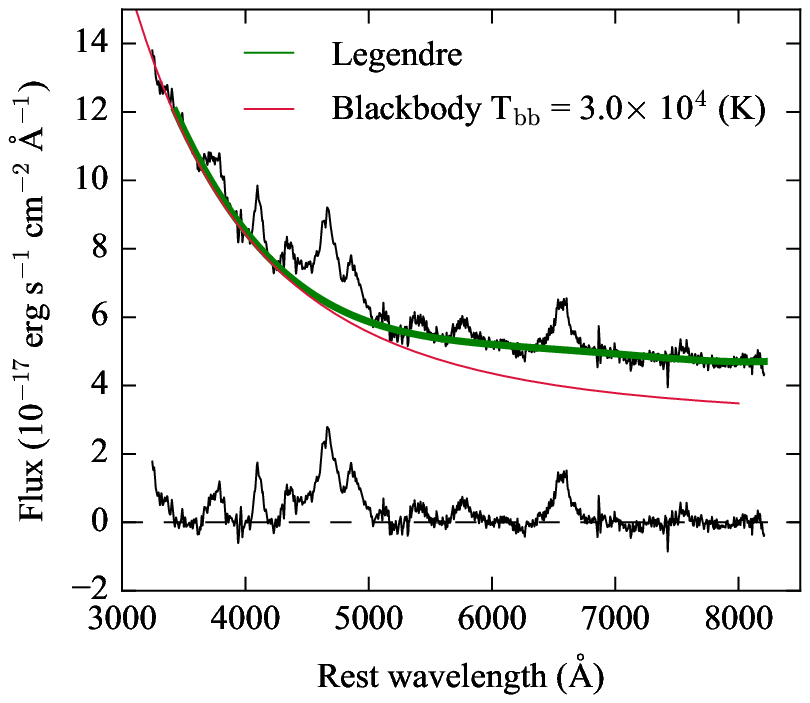}
\caption{An example of continuum subtraction of the host-subtracted LRIS spectrum from Jul 06. The black solid line shows the spectrum smoothed by 2 pixels. The green line shows the best-fit 5th-order Legendre polynomial while the red curve shows a blackbody spectrum. The spectrum near the dashed line is the residual from the subtraction of a 5th-order Legendre polynomial, which fits the spectrum better at shorter wavelengths.}
\label{fig:contfit}
\end{figure}

It is known that the spectroscopic signatures of an optically discovered TDEs consist of a strong blue continuum and a combination of broad HeII and H$\alpha$ emission \citep{Arcavi2014}. We select the regions outside of the Balmer lines and the He II emission line in the host subtracted spectrum to estimate the continuum. The line-free regions are fit with a 5th-order Legendre polynomial. A blackbody spectrum with $T_{bb}$ = 3.0$\times$ 10$^4$ (K) is shown in red in \autoref{fig:contfit}, which is the mean blackbody temperature from the SED fit. The blackbody spectrum shows depature from the host-subtracted spectrum at rest wavelength $\lambda > $ 4500 \AA.

\begin{figure*}
\centering
\includegraphics[width=7.0in, angle=0]{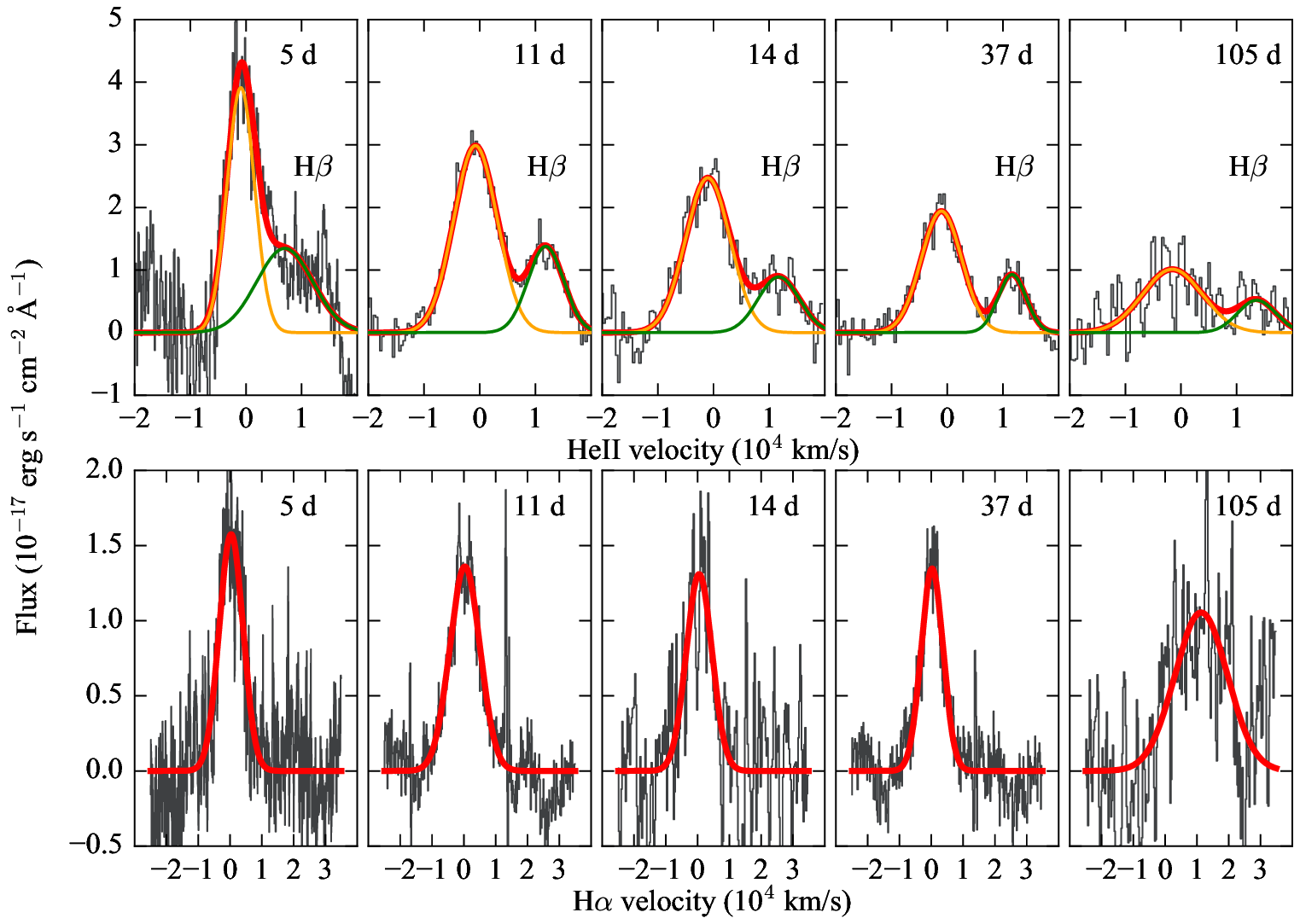}
\caption{Time evolution of He II and H$\alpha$ line profiles. The black solid lines show the TDE spectra after host and continuum subtraction. The TDE spectra are binned by a factor of 3 for clarity. He II (orange) and H$\beta$ (green) lines are fit simultaneously with two gaussian profiles to resolve spectral blending. The best-fit results are shown by the red solid lines. The time in the upper right corner corresponds to the time elapsed since discovery (MJD \tdisc{}).}
\label{fig:linefit}
\end{figure*}

We approximate the continuum with the Legendre polynomial because it fits the spectra better than the blackbody spectrum and does not require an assumption of the physical origin for the continuum. The line profiles of \HeII{} and H$\alpha$ are measured after host and TDE continuum subtraction. The best-fit results are shown as red lines in \autoref{fig:linefit}, where the grey solid line is the flux of the subtracted spectrum centered at the indicated line in velocity space. We simultaneously fit the \HeII{} (orange) and H$\beta$ (green) emissions as two individual Gaussian profiles. The H$\alpha$ line is modelled as a single Gaussian. The linewidths and line luminosities are listed in \autoref{table:emit}.

\section{Discussion}
\label{sec:discussion}
In this section, we start with discussing other potential mechanisms that could drive the observed flare and the implication of the derived rise time for iPTF16axa. We then compare its properties with 11 TDE candidates discovered in UV and optical sky surveys with well-sampled optical light curves: D1-9 and D3-13 from \textsl{GALEX}+CFHTLS \citep{Gezari2008}, TDE1 and TDE2 from SDSS \citep{vanVelzen2011}, PTF09ge from PTF \citep{Arcavi2014}, PS1-10jh \citep{Gezari2012}, and PS1-11af \citep{Chornock2014} from \textsl{GALEX}+Pan-STARRS1, ASASSN-14ae \citep{Holoien2014}, ASASSN-14li \citep{Holoien2016}, and ASASSN-15oi \citep{Holoien2016b} from ASASSN, and iPTF 16fnl \citep{2017arXiv170300965B}. The luminosities, temperatures, and radii of the three ASASSN candidate TDEs (ASASSN-14ae, ASASSN-14li, ASASSN-15oi) are provided by T. Holoien via private communication. We calculate the luminosities and radii for the other TDE candidates by scaling the best-sampled optical light curve ($g$ or $r$-band) to the peak bolometric luminosity reported in the literature, and assuming a constant temperature fixed to the value reported in the literature.

\subsection{Origin of the Flare}
The color evolution and the spectroscopic signatures of the flare are not consistent with any known supernova. Supernovae exhibit a much faster color evolution due to cooling in the expanding ejecta and can only remain bright in the UV for a few days. In addition, we do not detect any P-Cygni profile indicative of outflow in the spectra of iPTF16axa.

Although the nuclear position of the flare may connect it to AGN activity, we do not see any evidence of the host galaxy harbouring an active nucleus. Firstly, common AGN lines such as [OIII] and [NII] are not present in the spectra. Although the Balmer lines H$\alpha$ and H$\beta$ were detected, the broad Balmer lines have faded almost entirely from June 2016 to September 2016, which indicates the presence of broad Balmer lines is associated with the transient instead of the host galaxy. In fact, in the rare case of a changing-look AGN, we may see broad emission lines in an AGN suddenly appear or disappear on the timescale of a few years \citep[e.g.][]{2015ApJ...800..144L,2017ApJ...835..144G,2014ApJ...788...48S,2016ApJ...826..188R}. However, the lack of X-ray emission in iPTF16axa does not support the changing-look AGN scenario. Furthermore, AGN are known to vary on various timescales across the electromagnetic spectrum. As mentioned in \autoref{sec:discovery}, the position of iPTF16axa does not have any historical PTF detection signposting AGN activity between 2011 and 2014.

The photometric and spectroscopic properties bear a stronger resemblance to previous events classified as optical TDE candidates. We compare and discuss their temperatures, luminosities, and spectral line ratios in \autoref{subsec:discuss_temperature},\autoref{subsec:bol_lum}, and \autoref{subsec:line_ratio}.

\subsection{Timescale}
We derive the shortest rise time ($t_0$-$t_D$) by setting the derivative of Eq. A2 (in \citealt{Guillochon2013}) with respect to the impact parameter $\beta$ to zero. The minimum theoretical timescale implied by a 5$\times$10$^6$ \Msun{} black hole is 63 days for $\beta$=1.9 assuming a $\gamma$=4/3 and a solar type star. Since the TDE was discovered on the decline, we can only place an upper limit on the rise time derived from the observed light curve. The upper limit on the rise time is $\Delta t<$\restdt{} rest frame days assuming the peak light occurred some time before the iPTF discovery. This rise time is consistent with a black hole mass of less than 3$\times$10$^6$ \Msun{}, which is within the intrinsic scatter (0.38 dex) of the M-$\sigma$ relation in \cite{2013ApJ...764..184M} given in \autoref{subsec:host_prop}.

\subsection{Temperatures}
\label{subsec:discuss_temperature}
The blackbody temperature of iPTF16axa remained constant ($T_{bb}$$\sim$3.0$\times$10$^4$ K) over the 3 month monitoring period. This temperature is similar to what was found in PS1-10jh \citep{Gezari2012}, which was also reported to have constant temperature on the timescale of about a year. 

\begin{figure}[ht]
\centering
\includegraphics[width=3.5in, angle=0]{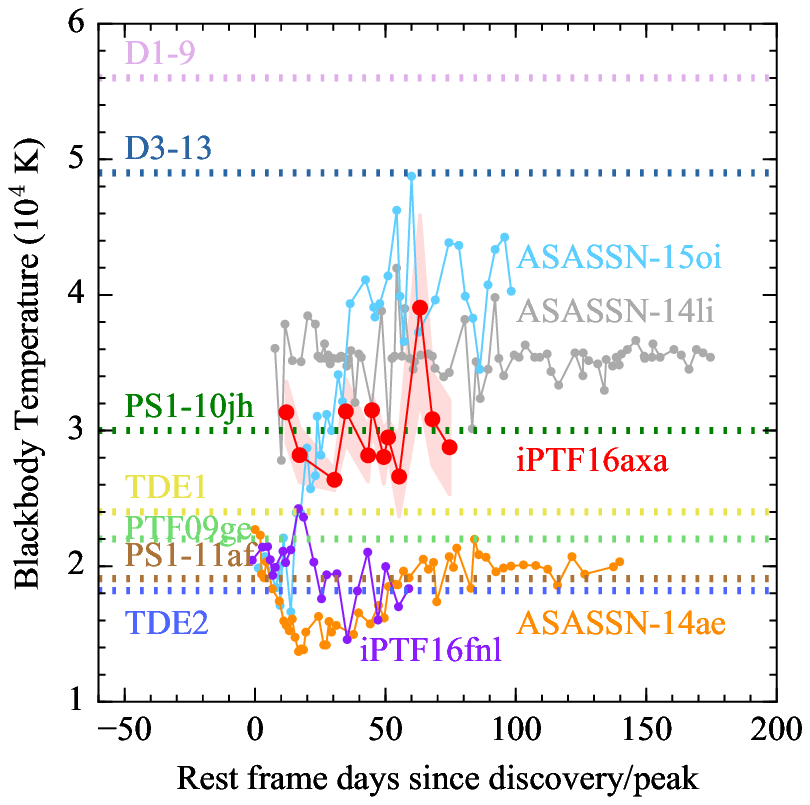}
\caption{Comparison of the evolution of the blackbody temperature inferred from SED fitting. The blackbody temperatures of the UV/optical TDE candidates remain constant on the order of a few 10$^4$K over time except ASASSN-15oi.}
\label{fig:tempcomp}
\end{figure}

The TDE candidates discovered by \textsl{GALEX} (D1-9, D3-13), which were also detected in the optical with CFHTLS, have higher blackbody temperatures than the other optical TDE candidates in \autoref{fig:tempcomp}. However, the difference is much less significant than the difference between X-ray-detected TDE candidates and optical TDE candidates, where the former is usually 1--2 orders of magnitude hotter than the latter.

TDE candidates found in the All-Sky Automated Survey for Supernovae (ASASSN), ASASSN-14ae and ASASSN-14li, the SDSS TDEs TDE1 and TDE2, PS1-10jh, PS1-11af, and PTF09ge also have blackbody temperatures that remain roughly constant over months (\autoref{fig:tempcomp}). The only outlier here is ASASSN-15oi, which features a $\sim$100\% increase in blackbody temperature on the timescale of less than a month.

\subsection{Helium-to-Hydrogen Ratio}
\label{subsec:line_ratio}

\begin{figure}[ht]
\centering
\includegraphics[width=3.5in, angle=0]{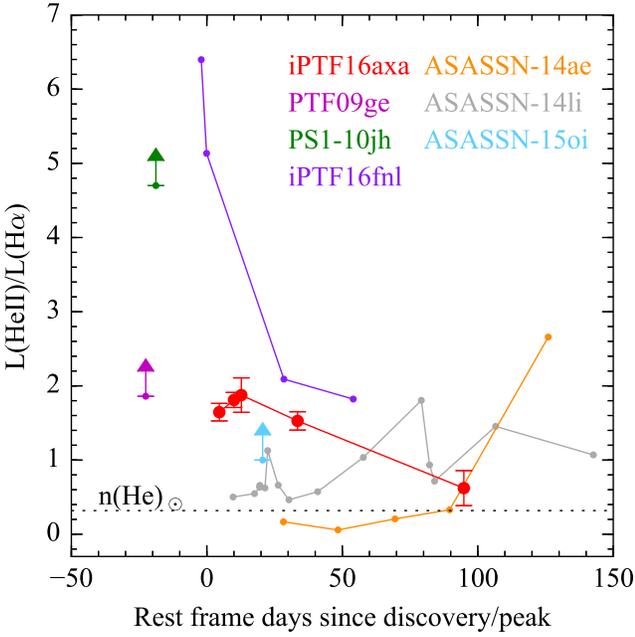}
\caption{Comparison of the evolution of the helium-to-hydrogen line ratio inferred from spectral fitting. The x-axis shows the time elapsed since peak (t$_0$) for PTF09ge, PS1-10jh, and iPTF16fnl, and time elapsed since \emph{discovery} for the ASASSN TDEs and iPTF16axa (MJD \tdisc{}). The dotted line shows the expected helium-to-hydrogen ratio in a nebular environment assuming the solar abundance of He/H. It is noticed that nebular arguments may not be valid for TDE candidates despite being frequently used in literature.}
\label{fig:ratio}
\end{figure}

\autoref{fig:ratio} shows the integrated Helium-to-H$\alpha$ line ratio of iPTF16axa and the other TDE candidates discovered in the optical. PS1-10jh, PTF09ge, and ASASSN-15oi do not have H$\alpha$ emission. A lower limit of 4.7 was reported for PS1-10jh \citep{2015ApJ...815L...5G}, a lower limit of $\sim 1$ was reported for ASASSN-15oi \citep{Holoien2016b}, and we measure a lower limit of 1.9 for PTF09ge from fitting its spectrum obtained from the Double Spectrograph mounted on the Palomar 200-inch (P200) telescope on 2009 May 20. 

We measure the line ratios for ASASSN-14ae and ASASSN-14li by performing Gaussian line fit on the spectra on the open TDE catalog \footnote{\url{tde.space}} in a similar fashion as described in \autoref{subsec:spectral}. The continuum is modelled as a 5th-order Legendre polynomial and subtracted before measuring the lines. ASASSN-14ae did not develop \HeII{} until later epochs.

Throughout the spectroscopic epochs, the H$\alpha$ line was readily detected in iPTF16axa except for the last epoch. iPTF16axa did not show significant H$\alpha$ suppression as was observed in PS1-10jh and PTF09ge. From \autoref{fig:ratio}, the spectroscopic signatures of TDE candidates can be divided into two groups based on the presence/absence of H$\alpha$ emission. The sources that show both \HeII{} and H$\alpha$ emission appear to have similar He II/Halpha ratios, with the exception of iPTF16fnl near peak, which shows a high \HeII-to-H$\alpha$ ratio that rapidly evolves to the lower ratio observed in the other sources.

The nebular \HeII{} to H$\alpha$ line ratio can be expressed as
\begin{equation}
\label{eq:lineratio}
\frac{L (\textrm{\HeII})}{L ( \mathrm{H\alpha} )} = \frac{n(He^{++}) n_e \alpha_{\lambda \textrm{4686}}^{eff}h\nu_{\lambda\textrm{4686}}}{n_p n_e \alpha_{\lambda \mathrm{H\beta}}^{eff}(j_{H\alpha}/j_{H\beta})h\nu_{\lambda\mathrm{H\beta}}},
\end{equation}
where $n(He^{++})$ is the density of He$^{++}$, $n_p$ is the proton density, $n_e$ is the electron density, and $\alpha_{\lambda}^{eff}$ is the effective recombination coefficient. For a typical T=10$^4$~K nebular gas, $\alpha_{\lambda4686}^{eff}$ = 3.57$\times$ 10$^{-13}$ cm$^3$ s$^{-1}$, $\alpha_{H_\beta}^{eff}$ = 3.02$\times$ 10$^{-14}$ cm$^3$ s$^{-1}$, and j$_{H\alpha}$/j$_{H\beta}$ is 2.87 (Osterbrock p80). Substituting in these values, the \HeII{} to H$\alpha$ line ratio can be expressed as 3.98 $n(He^{++})/n_{p}$ for an electron density of 10$^2$ cm$^{-3}$ in case B recombination. Assuming the solar helium abundance Y$_{\odot}$ = 0.2485 \citep{2010ApJ...719..865S}, the number abundance of helium $n(He^{++})/n_{p}$ is $\approx$0.08. This results in a line ratio of 0.32($\frac{n_{He}}{n_{He,\odot}}$), which is denoted by the dotted line in \autoref{fig:ratio}.

It is noticed that the nebular arguments, while still commonly used in the literature, are not valid for most of the TDE spectra. \autoref{fig:ratio} demonstrates that all measurements of the helium-to-hydrogen line ratio in TDEs, with the exception of the early epochs of ASASSN-14ae, display a helium enhancement compared to the nebular prediction assuming solar abundance. While stellar composition may be affecting these ratios in some events, this pattern also suggests that nebular arguments along the lines of \autoref{eq:lineratio} may break down for TDEs. A likely explanation is that high gas densities ($> 10^{10}$ cm$^{-3}$) are leading to the suppression of the Balmer lines as these transitions become optically thick. This possibility was first suggested by \cite{2004ApJ...610..707B}, and has been recently studied with CLOUDY caclulations \citep{2014MNRAS.438L..36G,2016arXiv161208093S,2015MNRAS.454.2321S} and full radiative transfer calculations \citep{Roth2016}. 

\subsection{Bolometric Luminosity}
\label{subsec:bol_lum}
\begin{figure}[ht]
\centering
\includegraphics[width=3.5in, angle=0]{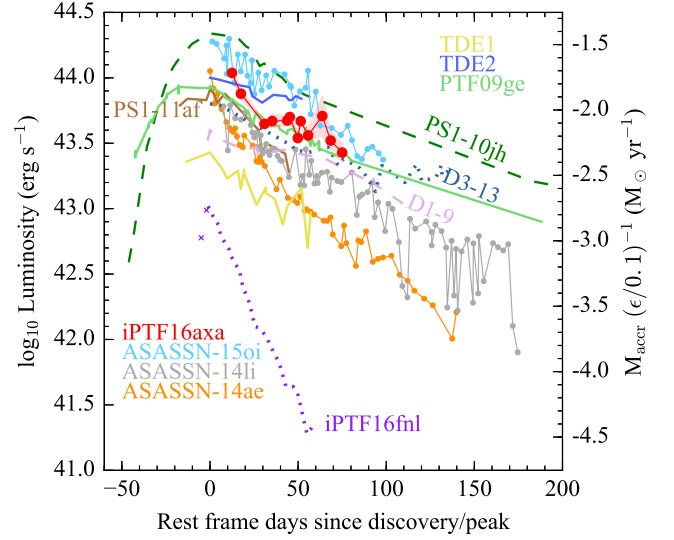}
\caption{Comparison of the evolution of the integrated UV-optical luminosity inferred from SED fitting. The y-axis on the right hand side is the mass accretion rate assuming an efficiency of 0.1. The x-axis shows the time elapsed since peak (t$_0$) for PTF09ge, PS1-10jh, and iPTF16fnl and the time elapsed since discovery for the ASASSN TDEs and iPTF16axa (MJD \tdisc{}). The two crosses in purple are derived from pre-peak $g$ band data of iPTF16fnl assuming a blackbody temperature of 2$\times$10$^4$K. It is worth noting that all of the UV and optically detected TDE candidates discussed here follow a t$^{-5/3} $power law decay except iPTF16fnl. These TDE candidates span a narrow range in the peak luminosity log(L [erg s$^{-1}$]) = 43.4 -- 44.4.}
\label{fig:lumcomp}
\end{figure}

Shown in \autoref{fig:lumcomp} is the time evolution of the UV-optical integrated luminosity of iPTF16axa from the blackbody model. Also shown in this plot are the UV/optical integrated luminosities of ASASSN-14ae, ASASSN-14li, ASASSN-15oi, PS1-10jh, PS1-11af, TDE1, TDE2, D1-9, and D3-13.

In \autoref{fig:lumcomp}, all of the TDE candidates except iPTF16fnl follow a power-law decline with a decline rate more or less consistent with $t^{-5/3}$. It is also interesting that, based on our blackbody fit, all of these TDEs except iPTF16fnl are confined to a small range of luminosities, with the peak luminosities ranging from log($L$ [erg s$^{-1}$])= $43.4- 44.4$. We must caution, however, that a substantial fraction of the total radiated energy, especially if originally emitted at FUV and EUV wavelengths, may be missing in our observations, as was demonstrated by \citet{2016ApJ...829...19V} in the case of PTF09ge based on infrared light echo observations.

\subsection{Photospheric Radius}
\begin{figure}
\centering
\includegraphics[width=3.5in, angle=0]{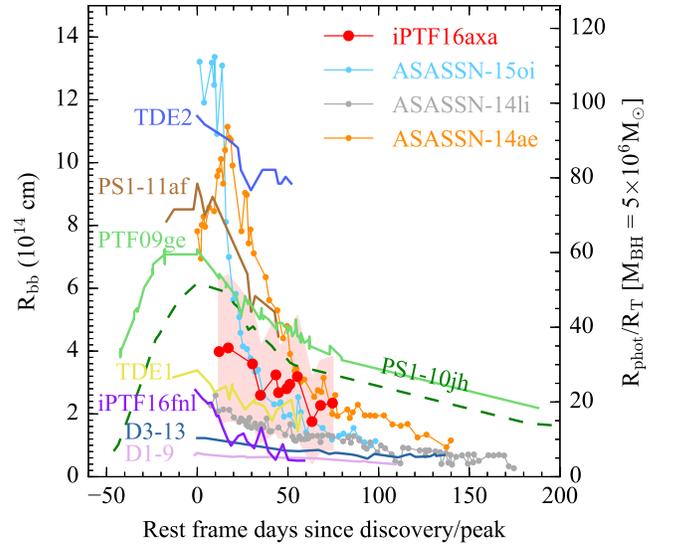}
\caption{Comparison of the evolution of the blackbody radius ($R_{bb}$) inferred from SED fitting. The dots in the figure represent $R_{bb}$ derived from the SED some time after discovery for iPTF16axa and the ASASSN objects. The pink shaded area shows the uncertainties of $R_{bb}$ for iPTF16axa. The blackbody radii derived are on the order of a few 10 times of the tidal radius.}
\label{fig:radiuscomp}
\end{figure}

\autoref{fig:radiuscomp} shows the evolution of blackbody radius for iPTF16axa and other optically bright TDE candidates. The blackbody radius of iPTF16axa decreased steadily from 4$\times$10$^{14}$ cm to 2$\times$10$^{14}$ cm as the luminosity decreases with time. The blackbody radius of PS1-10jh is derived assuming a $t^{-5/3}$ decay in luminosity and constant temperature. Since the tidal radius is weakly dependent on the black hole mass ($R_T \propto M_{\rm BH}^{1/3}$), \autoref{fig:radiuscomp} shows that the derived radii are at least 10 times farther away from the $R_T$ for all the TDE candidates.

Due to the non-varying temperature evolution of TDE emission, the photospheric radius must decline at late times in order to match the fading light curve. The physical meaning of this decline remains unclear. One explanation is that the density of the optically emitting gas drops over time, allowing the observer to see light emitted from increasingly deeper regions, even if the gas is continuously outflowing \citep{2015MNRAS.454.2321S}. Another possibility is that the optically emitting gas is in fact moving closer to the black hole over time, and may be related to the decreasing apocenter radius of the circularizing debris stream \citep{2017MNRAS.464.2816B}.

\subsection{Virial radius}

\begin{figure}[ht]
\centering
\includegraphics[width=3.5in, angle=0]{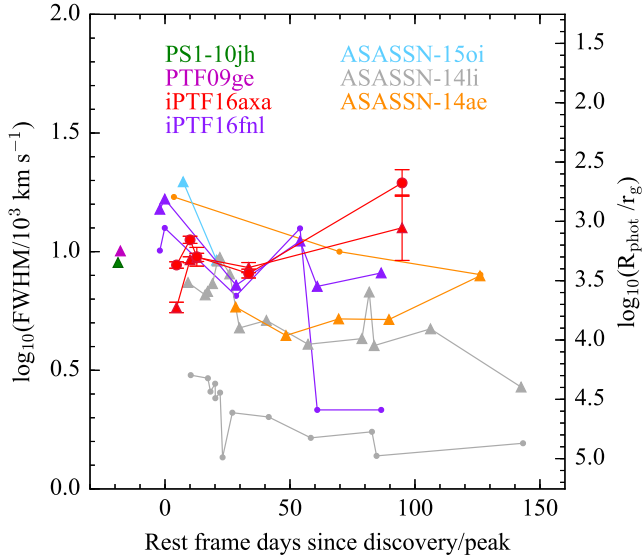}
\caption{Comparison of the evolution of the photosphere radius inferred from emission linewidths. The triangles mark the linewidths of \HeII{} lines while the dots mark the linewidths of H$\alpha$ emission.  The y-axis on the right hand side shows the photospheric radii in units of the gravitational radius $r_g = GM/c^2$. Throughout the monitoring period, the FWHM of H$\alpha$ and \HeII{} evolve in a similar trend. The fact that \HeII{} line is not wider than H$\alpha$ disfavors the scenario of a stratified BLR region that is virially bound.}
\label{fig:lineradius}
\end{figure}

The FWHMs of \HeII{} and H$\alpha$ are plotted in \autoref{fig:lineradius}. The triangles denote the linewidths of \HeII{} lines while the dots denote the linewidths of H$\alpha$ emission. 
The \HeII{} linewidths for ASASSN-14ae and ASASSN-14li were measured using the spectra on the open TDE catalog. The \HeII{} linewidth of PTF09ge is measured from its P200 spectrum and the value for PS1-10jh is provided in \cite{Gezari2012}.

In \autoref{fig:lineradius}, the FWHMs of He~II and H$\alpha$ emission lines evolve in the same trend. Throughout the observations of iPTF16axa, the \HeII{} linewidth remains comparable, sometimes even narrower, than the linewidths of H$\alpha$. The fact that the linewidths of He~II are not wider than that of H$\alpha$ suggests the line emitting material is not virially bound. In the scenario of a stratified broad line region, because the photoionization energy of He is higher than hydrogen, helium has to be emitted at a smaller radius and therefore would have a wider linewidth. As pointed out in \cite{Holoien2016,Holoien2016b}, in reverberation mapping studies, the linewidths would increase while the luminosity decreases due to recombination at outer radii. This trend is also not observed until the last epoch when the line detection was weak.

\begin{deluxetable*}{lcccc}
\tablewidth{0pc}
\tabletypesize{\scriptsize}
\tablecolumns{5}
\tablecaption{ Emission line fit \label{table:emit}} 
\centering
\tablehead{\colhead{Date} & \colhead{He II FWHM} & \colhead{L(He II)} & \colhead{H$\alpha$ FWHM} & \colhead{L(H$\alpha$)} \\
 & \colhead{(10$^3$ km s$^{-1}$)} & \colhead{(10$^{40}$ erg s$^{-1}$)} & \colhead{(10$^3$ km s$^{-1}$)} & \colhead{(10$^{40}$ erg s$^{-1}$)}}
\startdata 
2016-06-04	&	 5.8$\pm$0.3	&	12.5$\pm$0.8 	&	8.8$\pm$0.3	&	7.6$\pm$0.3 \\
2016-06-10	&	 9.2$\pm$0.3	&	15.1$\pm$0.6	&	11.2$\pm$0.4	&	8.3$\pm$0.4 \\
2016-06-13	&	 9.5$\pm$0.4	&	12.8$\pm$0.7	&	9.5$\pm$0.9	&	6.8$\pm$0.8 \\
2016-07-06	&	 8.6$\pm$0.4	&	9.1$\pm$0.6	&	8.1$\pm$0.3	&	6.0$\pm$0.3 \\
2016-09-12	&	 12.6$\pm$4.2	&	7.0$\pm$2.6	&	19.5$\pm$2.4	&	11.2$\pm$1.5
\enddata

\end{deluxetable*}

\subsection{Peak luminosity}
\begin{figure}[ht]
\centering
\includegraphics[width=3.5in, angle=0]{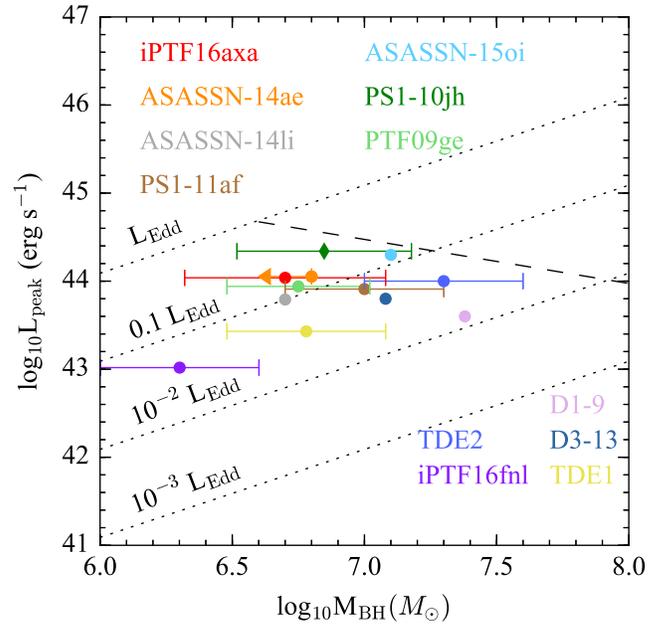}
\caption{The peak luminosities for TDEs from \autoref{fig:lumcomp} vs black hole masses. Black hole masses obtained from literature are marked in circles while triangles are black hole masses derived using r-band scaling in \cite{2007ApJ...663...53T}. The dotted lines show the luminosities that correspond to 4 different Eddington ratios while the black dashed line shows the $\dot{M}_{peak}$ $\propto$ $M_{\rm BH}^{-1/2}$ relation expected from theoretical work normalized to Eq. A1 in \cite{Guillochon2013} with $\gamma$=4/3, $\beta$=1, and $\epsilon$=0.1. Below $M_{\rm BH}\sim 10^{6.6}$ \Msun{}, the luminiosity should be Eddington-limited and scales proportionally with $M_{\rm BH}$.}
\label{fig:Lpeak}
\end{figure}

In \autoref{fig:Lpeak} we plot the peak luminosity reported in the literature as a function of the black hole mass. The circle symbols show black hole masses reported in literature while the diamond symbols show black hole masses estimated from the $r$-band scaling relation in \cite{2007ApJ...663...53T}, which has a 1$\sigma$ scatter of 0.33 dex. We obtain black hole mass of ASASSN-14ae from \cite{Holoien2014}, ASASSN-14li from \cite{Holoien2016}, ASASSN-15oi from \cite{Holoien2016b}, PS1-11af from \cite{Chornock2014}, PTF09ge from \cite{Arcavi2014}, D1-9 and D3-13 from \cite{2009ApJ...698.1367G}, and TDE1 and TDE2 from \cite{vanVelzen2011}.

We show four different ratios of Eddington luminosity, $L_{Edd}$, 0.1$L_{Edd}$, 10$^{-2}$$L_{Edd}$, 10$^{-3}$$L_{Edd}$, as a function of the black hole mass with the black dotted lines. The black dashed line in \autoref{fig:Lpeak} shows the theoretical scaling of $L_{peak}$~$\propto$~$\dot{M}_{peak}$~$\propto$ $M_{\rm BH}^{-1/2}$ \citep{2011MNRAS.410..359L,Guillochon2013} normalized to Eq. A1 in \cite{Guillochon2013} assuming a star with solar mass and radius, $\gamma$=4/3, $\beta$=1, and an accretion efficiency $\epsilon$ of 0.1. The dashed line does not extend below $M_{\rm BH}\sim 10^{6.6}$ \Msun{} since the emergent luminosity should be Eddington limited. Below this threshold, the luminosity scales with the Eddington luminosity, $L_{peak}$~$\propto$~$L_{Edd}~\propto$~$M_{\rm BH}$. We do not see a clear trend in the data that suggests the peak luminosity and the black hole mass are correlated, although we emphasize again that undetected emission originally at FUV and EUV wavelengths may alter this conclusion, and that many of the TDE candidates plotted were discovered post-peak, and thus their luminosity at discovery may be underestimating the true peak luminosity.

\section{Conclusion}
We present results from photometric and spectroscopic follow up observations of a strong TDE candidate, iPTF16axa, and comparisons of the derived physical quantities with 11 other optically studied TDE candidates from ASASSN, GALEX+CFHTLS, PTF, PS1, and SDSS. Both \Swift{} UVOT observations and the follow up spectra of iPTF16axa are consistent with the object being a TDE rather than a supernova or a variable AGN. The UV and optical light curves of iPTF16axa are in good agreement with the $t^{-5/3}$ relation and suggest the TDE was discovered \restdt{} rest-frame days after disruption. The light curve shows no color evolution with time, with an SED fitted with a constant temperature of 3$\times$10$^4$~K. The TDE is hosted by an early-type galaxy with an estimated black hole mass of 5$\times$10$^6$\Msun{}, which is similar to previously reported TDE candidate hosts. We summarize the comparisons of a sample totaling 12 TDE candidates including iPTF16axa below.

\begin{enumerate}
\item TDE candidates discovered in the UV and optical remain roughly constant temperature over several months. The blackbody temperatures of the TDE candidates are found to be a few 10$^4$~K.
\item Our sample of TDE candidates are characterized by a power law decline and, based on a blackbody fit to optical and near-UV data, span a small range of peak luminosity of 10$^{43.4}<$$L_{peak}$$<$10$^{44.4}$. The decline is more or less consistent with the classic $t^{-5/3}$ prediction except in iPTF16fnl, which fades more steeply than the other TDE candidates discussed in this paper.
\item Nebular arguments are not valid for interpreting line ratios in most optically discovered TDE candidates due to the presence of high density gas, which can lead to the suppression of hydrogen Balmer transitions. The spectra of UV/optical TDE candidates show a range of He-to-H$\alpha$ ratios, and the time-evolution of these ratios also differs between events. Detailed modeling will be necessary to understand these behaviors.
\item The blackbody radii derived from the SEDs of UV/optical TDE candidates trace distances that are much larger than the tidal radius ($\approx$ a few 10R$_T$), and that decline with time.
\item The FWHM of \HeII{} is consistent with the FWHM of H$\alpha$ in the optical spectra of UV/optical TDE candidates. This evidence contradicts the assumption of a stratified BLR.
\item Theoretical work shows that the peak luminosity and the black hole mass are correlated by $L_{peak}$ $\propto$ $M_{\rm BH}^{-1/2}$ except at smaller black hole masses, where the emission is Eddington-capped ($L_{peak}$ $\propto$ $M_{\rm BH}$). However, there is no strong trend between the two quantities in the sample of candidate TDEs discovered in UV and optical.
\end{enumerate}

\section{Acknowledgements}
We thank the anonymous referee for their helpful comments on the manuscript. T.H. thanks T. Holoien and C. Bonnerot for providing data from their papers.  S.G. is supported in part by NSF CAREER grant 1454816, NASA \Swift{} Cycle 12 grant NNX16AN85G, and NASA Keck Grant 1568615. N.R. acknowledges the support of a Joint Space-Science Institute prize postdoctoral fellowship. A.H. acknowledges support by a grant from the I-CORE program "From the Big Bang to Planets". Support for I.A. was provided by NASA through the Einstein Fellowship Program, grant PF6-170148. These results made use of the Discovery Channel Telescope  at  Lowell  Observatory. Lowell  is  a  private,  non-profit institution dedicated to astrophysical research and public appreciation of astronomy and operates the DCT in partnership with Boston University, the University of Maryland, the University of Toledo, Northern Arizona University and Yale University.
The W. M. Keck Observatory is operated as a scientific partnership among the California Institute of Technology, the University of California, and NASA; the Observatory was made possible by the generous financial support of the W. M. Keck Foundation.  This research used resources of the National Energy Research Scientific Computing Center, a DOE Office of Science User Facility supported by the Office of Science of the U.S. Department of Energy under Contract No. DE-AC02-05CH11231.  
\clearpage
\bibliography{tde}

\clearpage
\appendix

\setcounter{table}{0}
\renewcommand{\thetable}{A\arabic{table}}
\LongTables
\begin{deluxetable}{lccc}

\tablecolumns{4}
\tablecaption{Photometric data of iPTF16axa} 
\centering
\tablehead{\colhead{MJD} & \colhead{Magnitude} & \colhead{Filter} & \colhead{Telescope}}

\startdata 
57546.189 & 19.110 $\pm$ 0.050 & UVW2 & \Swift{} \\
57551.232 & 19.070 $\pm$ 0.060 & UVW2 & \Swift{} \\
57556.824 & 19.350 $\pm$ 0.070 & UVW2 & \Swift{} \\
57571.634 & 19.920 $\pm$ 0.090 & UVW2 & \Swift{} \\
57576.485 & 20.040 $\pm$ 0.100 & UVW2 & \Swift{} \\
57585.937 & 19.920 $\pm$ 0.110 & UVW2 & \Swift{} \\
57592.583 & 20.220 $\pm$ 0.160 & UVW2 & \Swift{} \\
57594.428 & 19.870 $\pm$ 0.060 & UVW2 & \Swift{} \\
57599.095 & 20.120 $\pm$ 0.080 & UVW2 & \Swift{} \\
57607.999 & 20.300 $\pm$ 0.080 & UVW2 & \Swift{} \\
57613.241 & 20.380 $\pm$ 0.070 & UVW2 & \Swift{} \\
57620.542 & 20.470 $\pm$ 0.090 & UVW2 & \Swift{} \\
57551.236 & 19.250 $\pm$ 0.070 & UVM2 & \Swift{} \\
57556.828 & 19.500 $\pm$ 0.070 & UVM2 & \Swift{} \\
57571.637 & 19.980 $\pm$ 0.090 & UVM2 & \Swift{} \\
57576.487 & 20.070 $\pm$ 0.100 & UVM2 & \Swift{} \\
57587.593 & 20.110 $\pm$ 0.110 & UVM2 & \Swift{} \\
57592.586 & 20.450 $\pm$ 0.250 & UVM2 & \Swift{} \\
57594.436 & 20.060 $\pm$ 0.100 & UVM2 & \Swift{} \\
57599.099 & 20.160 $\pm$ 0.160 & UVM2 & \Swift{} \\
57608.002 & 20.640 $\pm$ 0.130 & UVM2 & \Swift{} \\
57613.248 & 20.550 $\pm$ 0.120 & UVM2 & \Swift{} \\
57620.546 & 20.740 $\pm$ 0.150 & UVM2 & \Swift{} \\
57551.230 & 19.320 $\pm$ 0.090 & UVW1 & \Swift{} \\
57556.820 & 19.500 $\pm$ 0.100 & UVW1 & \Swift{} \\
57561.391 & 19.650 $\pm$ 0.160 & UVW1 & \Swift{} \\
57571.630 & 19.780 $\pm$ 0.120 & UVW1 & \Swift{} \\
57576.483 & 20.320 $\pm$ 0.180 & UVW1 & \Swift{} \\
57585.933 & 19.910 $\pm$ 0.170 & UVW1 & \Swift{} \\
57592.581 & 20.020 $\pm$ 0.230 & UVW1 & \Swift{} \\
57594.422 & 20.500 $\pm$ 0.140 & UVW1 & \Swift{} \\
57599.092 & 20.190 $\pm$ 0.130 & UVW1 & \Swift{} \\
57607.997 & 20.430 $\pm$ 0.140 & UVW1 & \Swift{} \\
57613.235 & 20.430 $\pm$ 0.130 & UVW1 & \Swift{} \\
57620.538 & 20.740 $\pm$ 0.190 & UVW1 & \Swift{} \\
57551.231 & 19.300 $\pm$ 0.140 & UVOT-U & \Swift{} \\
57556.822 & 19.410 $\pm$ 0.150 & UVOT-U & \Swift{} \\
57571.632 & 19.950 $\pm$ 0.220 & UVOT-U & \Swift{} \\
57576.484 & 20.090 $\pm$ 0.250 & UVOT-U & \Swift{} \\
57585.935 & 19.820 $\pm$ 0.220 & UVOT-U & \Swift{} \\
57587.587 & 19.820 $\pm$ 0.130 & UVOT-U & \Swift{} \\
57594.425 & 19.940 $\pm$ 0.130 & UVOT-U & \Swift{} \\
57599.093 & 20.220 $\pm$ 0.190 & UVOT-U & \Swift{} \\
57607.998 & 20.170 $\pm$ 0.170 & UVOT-U & \Swift{} \\
57613.238 & 20.220 $\pm$ 0.150 & UVOT-U & \Swift{} \\
57620.540 & 20.140 $\pm$ 0.200 & UVOT-U & \Swift{} \\
57537.397 & 19.486 $\pm$ 0.071 & g & P48 \\
57540.404 & 19.615 $\pm$ 0.032 & g & P60 \\
57541.428 & 19.592 $\pm$ 0.031 & g & P60 \\
57544.370 & 19.464 $\pm$ 0.081 & g & P48 \\
57547.399 & 19.724 $\pm$ 0.027 & g & P60 \\
57548.385 & 19.626 $\pm$ 0.085 & g & P48 \\
57548.510 & 20.119 $\pm$ 0.137 & gp & LCO \\
57552.373 & 19.673 $\pm$ 0.103 & g & P48 \\
57554.378 & 19.959 $\pm$ 0.042 & g & P60 \\
57555.590 & 20.579 $\pm$ 0.157 & gp & LCO \\
57558.469 & 20.192 $\pm$ 0.127 & g & P60 \\
57565.219 & 20.165 $\pm$ 0.076 & g & P60 \\
57566.220 & 20.284 $\pm$ 0.027 & g & P60 \\
57574.502 & 20.918 $\pm$ 0.108 & gp & LCO \\
57581.465 & 20.931 $\pm$ 0.122 & gp & LCO \\
57583.261 & 20.611 $\pm$ 0.053 & g & P60 \\
57585.194 & 20.749 $\pm$ 0.070 & g & P60 \\
57587.193 & 20.699 $\pm$ 0.105 & g & P60 \\
57587.472 & 20.612 $\pm$ 0.274 & gp & LCO \\
57593.234 & 20.736 $\pm$ 0.048 & g & P60 \\
57595.182 & 20.771 $\pm$ 0.048 & g & P60 \\
57595.453 & 21.310 $\pm$ 0.278 & gp & LCO \\
57597.239 & 20.883 $\pm$ 0.056 & g & P60 \\
57599.197 & 20.875 $\pm$ 0.046 & g & P60 \\
57601.183 & 20.947 $\pm$ 0.081 & g & P60 \\
57605.213 & 21.041 $\pm$ 0.048 & g & P60 \\
57607.176 & 20.995 $\pm$ 0.052 & g & P60 \\
57609.177 & 21.011 $\pm$ 0.057 & g & P60 \\
57611.196 & 21.079 $\pm$ 0.067 & g & P60 \\
57613.173 & 21.038 $\pm$ 0.100 & g & P60 \\
57614.180 & 21.261 $\pm$ 0.141 & g & P60 \\
57615.168 & 21.221 $\pm$ 0.177 & g & P60 \\
57616.166 & 21.111 $\pm$ 0.113 & g & P60 \\
57617.169 & 20.903 $\pm$ 0.181 & g & P60 \\
57618.150 & 21.329 $\pm$ 0.191 & g & P60 \\
57619.156 & 20.976 $\pm$ 0.143 & g & P60 \\
57621.150 & 21.160 $\pm$ 0.120 & g & P60 \\
57625.158 & 21.203 $\pm$ 0.049 & g & P60 \\
57631.152 & 21.218 $\pm$ 0.071 & g & P60 \\
57633.166 & 21.297 $\pm$ 0.063 & g & P60 \\
57635.135 & 21.356 $\pm$ 0.138 & g & P60 \\
57639.227 & 21.413 $\pm$ 0.093 & g & P60 \\
57646.190 & 21.328 $\pm$ 0.330 & g & P60 \\
57656.182 & 21.536 $\pm$ 0.277 & g & P60 \\
57540.397 & 20.045 $\pm$ 0.033 & r & P60 \\
57541.421 & 19.972 $\pm$ 0.054 & r & P60 \\
57547.392 & 20.090 $\pm$ 0.029 & r & P60 \\
57548.514 & 19.999 $\pm$ 0.391 & rp & LCO \\
57551.451 & 20.162 $\pm$ 0.030 & r & P60 \\
57554.368 & 20.235 $\pm$ 0.074 & r & P60 \\
57558.459 & 20.404 $\pm$ 0.106 & r & P60 \\
57559.483 & 20.401 $\pm$ 0.260 & r & P60 \\
57565.198 & 20.527 $\pm$ 0.127 & r & P60 \\
57566.205 & 20.661 $\pm$ 0.029 & r & P60 \\
57569.229 & 20.711 $\pm$ 0.040 & r & P60 \\
57574.512 & 20.543 $\pm$ 0.283 & rp & LCO \\
57581.474 & 20.671 $\pm$ 0.297 & rp & LCO \\
57583.255 & 21.020 $\pm$ 0.087 & r & P60 \\
57585.188 & 21.149 $\pm$ 0.113 & r & P60 \\
57587.188 & 21.025 $\pm$ 0.144 & r & P60 \\
57587.480 & 20.580 $\pm$ 0.339 & rp & LCO \\
57593.228 & 21.063 $\pm$ 0.068 & r & P60 \\
57595.176 & 21.241 $\pm$ 0.102 & r & P60 \\
57595.462 & 20.603 $\pm$ 0.342 & rp & LCO \\
57597.234 & 21.213 $\pm$ 0.100 & r & P60 \\
57599.191 & 21.086 $\pm$ 0.077 & r & P60 \\
57605.207 & 21.280 $\pm$ 0.069 & r & P60 \\
57607.170 & 21.380 $\pm$ 0.105 & r & P60 \\
57609.171 & 21.317 $\pm$ 0.124 & r & P60 \\
57611.191 & 21.267 $\pm$ 0.108 & r & P60 \\
57613.167 & 21.377 $\pm$ 0.150 & r & P60 \\
57614.175 & 21.320 $\pm$ 0.171 & r & P60 \\
57616.161 & 21.284 $\pm$ 0.212 & r & P60 \\
57617.163 & 21.155 $\pm$ 0.244 & r & P60 \\
57621.144 & 21.398 $\pm$ 0.170 & r & P60 \\
57625.153 & 21.358 $\pm$ 0.107 & r & P60 \\
57629.148 & 21.478 $\pm$ 0.091 & r & P60 \\
57631.146 & 21.474 $\pm$ 0.160 & r & P60 \\
57633.161 & 21.464 $\pm$ 0.113 & r & P60 \\
57637.139 & 21.650 $\pm$ 0.226 & r & P60 \\
57639.222 & 21.634 $\pm$ 0.180 & r & P60 \\
57640.182 & 21.489 $\pm$ 0.096 & r & P60 \\
57646.151 & 21.640 $\pm$ 0.356 & r & P60 \\
57656.140 & 21.896 $\pm$ 0.527 & r & P60 \\
57540.401 & 19.901 $\pm$ 0.045 & i & P60 \\
57541.424 & 19.827 $\pm$ 0.050 & i & P60 \\
57547.395 & 19.891 $\pm$ 0.054 & i & P60 \\
57548.527 & 20.046 $\pm$ 0.378 & ip & LCO \\
57551.456 & 20.041 $\pm$ 0.052 & i & P60 \\
57554.373 & 20.063 $\pm$ 0.118 & i & P60 \\
57555.599 & 19.796 $\pm$ 0.357 & ip & LCO \\
57558.464 & 20.171 $\pm$ 0.143 & i & P60 \\
57565.202 & 20.408 $\pm$ 0.356 & i & P60 \\
57574.519 & 20.586 $\pm$ 0.311 & ip & LCO \\
57581.482 & 20.642 $\pm$ 0.326 & ip & LCO \\
57583.258 & 20.633 $\pm$ 0.127 & i & P60 \\
57585.191 & 20.615 $\pm$ 0.093 & i & P60 \\
57587.191 & 20.881 $\pm$ 0.144 & i & P60 \\
57587.489 & 20.537 $\pm$ 0.462 & ip & LCO \\
57593.231 & 20.829 $\pm$ 0.082 & i & P60 \\
57595.179 & 20.879 $\pm$ 0.143 & i & P60 \\
57595.470 & 20.747 $\pm$ 0.408 & ip & LCO \\
57597.237 & 20.769 $\pm$ 0.101 & i & P60 \\
57599.194 & 21.000 $\pm$ 0.101 & i & P60 \\
57605.210 & 21.116 $\pm$ 0.089 & i & P60 \\
57607.173 & 20.975 $\pm$ 0.125 & i & P60 \\
57609.174 & 21.109 $\pm$ 0.114 & i & P60 \\
57611.194 & 21.012 $\pm$ 0.223 & i & P60 \\
57613.170 & 20.974 $\pm$ 0.163 & i & P60 \\
57616.163 & 21.030 $\pm$ 0.189 & i & P60 \\
57617.166 & 21.165 $\pm$ 0.286 & i & P60 \\
57625.156 & 21.086 $\pm$ 0.125 & i & P60 \\
57631.149 & 21.146 $\pm$ 0.141 & i & P60 \\
57633.164 & 21.111 $\pm$ 0.155 & i & P60 \\
57639.224 & 21.417 $\pm$ 0.244 & i & P60 \\
57646.171 & 21.199 $\pm$ 0.256 & i & P60 \\
57656.163 & 21.213 $\pm$ 0.276 & i & P60 

\enddata
\end{deluxetable}
\end{document}